\makeatletter\@input{xx.tex}\makeatother
\documentclass[prl,superscriptaddress,twocolumn,noshowpacs]{revtex4-1}
\pdfoutput=1
\usepackage[pdftex]{graphicx}
\usepackage{amsmath}
\usepackage{filecontents}

\usepackage{amssymb}
\usepackage{mathtools} 
\usepackage{extarrows}
\usepackage{xr-hyper} 
\usepackage{hyperref}
\usepackage{epstopdf}
\usepackage{tikz}
\usepackage{capt-of}
\usepackage{tabularx}

\usepackage{times}
\usepackage{enumitem}
\usepackage{comment}
\usepackage{xfrac}
\usepackage{braket}
\usepackage{dsfont}

\DeclareMathOperator{\Tr}{Tr}

\newcommand{\centerfloat}{%
  \parindent \z@
  \leftskip \z@ \@plus 1fil \@minus \textwidth
  \rightskip\leftskip
  \parfillskip \z@skip}

\tikzset{dangerous style/.code={
    \tikzoption{clip}[]{\pgf@relevantforpicturesizefalse}
    \tikzoption{use as bounding box}[]{\pgf@relevantforpicturesizefalse}
    }
}

 \definecolor{blue}{HTML}{00538A} 
  \definecolor{red}{HTML}{C10020} 
  \definecolor{SkyBlue}{HTML}{CEA262}
\definecolor{orange}{HTML}{FF6800}

\begin{document}

\title{Symmetry enhanced boundary qubits at infinite temperature}

\author{Jack Kemp}
\affiliation{Rudolf Peierls Centre for Theoretical Physics, University of Oxford, 1 Keble Road, Oxford, OX1 3NP, UK}
\affiliation{Department of Physics, University of California, Berkeley, CA 94720, USA}
\author{Norman Y. Yao}
\affiliation{Department of Physics, University of California, Berkeley, CA 94720, USA}
\author{Chris R. Laumann}
\affiliation{Department of Physics, Boston University, Boston, MA, 02215, USA}


\begin{abstract}
 The $\mathbb{Z}_2 \times \mathbb{Z}_2$ symmetry protected topological (SPT) phase hosts a robust boundary qubit at zero temperature. At finite energy density, the SPT phase is destroyed and bulk observables equilibrate in finite time. Nevertheless, we predict parametric regimes in which the boundary qubit survives to arbitrarily high temperature, with an exponentially longer coherence time than that of the thermal bulk degrees of freedom. In a dual picture, the persistence of the qubit stems from the inability of the bulk to absorb the virtual $\mathbb{Z}_2 \times \mathbb{Z}_2$ domain walls emitted by the edge during the relaxation process. We confirm the long coherence time by exact diagonalization and connect it to the presence of a pair of conjugate almost strong zero modes. Our results provide a route to experimentally construct long-lived coherent boundary qubits at infinite temperature in disorder-free systems.

\end{abstract}

\maketitle

The primary signature of symmetry protected topological (SPT) order is the presence of robust boundary degrees of freedom at zero temperature. 
At \emph{finite} temperature, these boundary modes interact strongly with thermal excitations in the bulk and rapidly decohere. 
Recent progress on understanding many-body localized (MBL) states of matter \cite{nandkishore2015many,abanin2017recent} has yielded the insight that such edge modes can, in fact, be stabilized at finite temperature via strong quenched disorder~\cite{Chandran14, Potter15, Bahri15, Yao15}. 
In this case, the disorder serves to localize bulk thermal excitations, thereby preventing them from scattering with and decohering the boundary mode. 
While intriguing, the requirement of strong disorder complicates prospects for realizing such MBL SPT phases in experiments \cite{schreiber2015observation,choi2016exploring,smith2016many,luschen2017signatures,bordia2017periodically} and also weakens the distinguishing feature of the decoupled boundary mode, since bulk transport is also arrested.

In this Letter, we describe how the boundary and bulk degrees of freedom in a translationally invariant system can decouple parametrically, even at infinite temperature. This separation of edge and bulk dynamics stems from the inability of the edge to resonantly absorb or emit bulk excitations. We exploit this dynamical protection to construct a coherent edge qubit in a one-dimensional spin chain without disorder.

\begin{figure*}
 \includegraphics[width=0.9\linewidth]{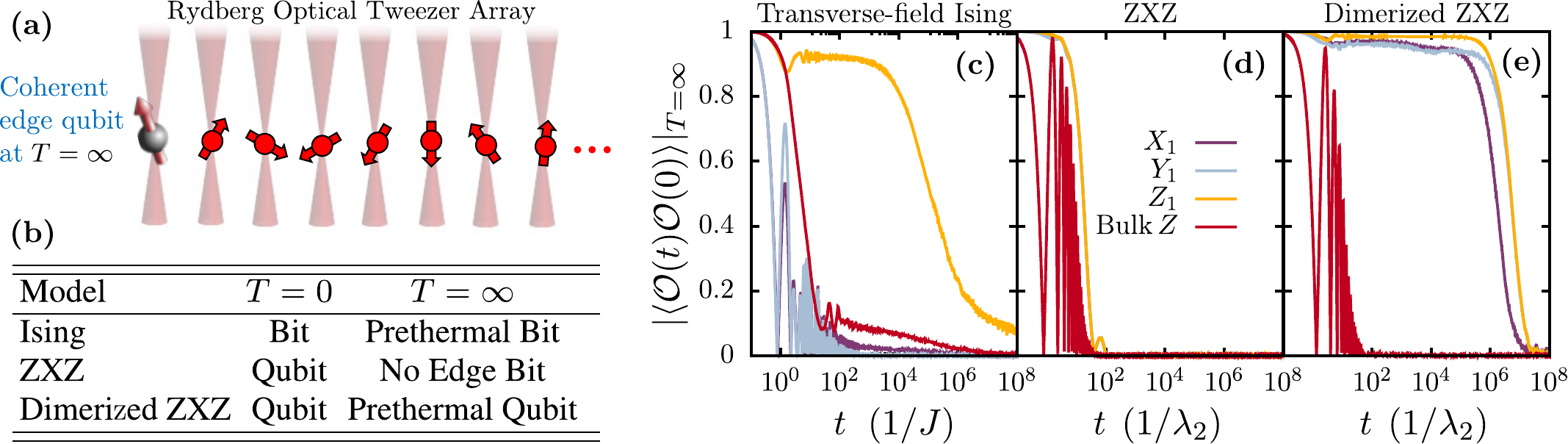}
\caption{%
(a) Schematic illustration of a one dimensional Rydberg optical tweezer array which hosts a coherent edge qubit even at infinite temperature. 
(b) Comparison of edge mode behavior at zero and infinite temperature in each of the models discussed in the main text.
(c-e) 
The auto-correlator of the edge spin operators at infinite temperature from exact diagonalisation for (c) the transverse-field Ising chain of Eq.~\eqref{eq:IsingHam} with $J_2$ = $\Gamma = 0.25 J$, compared with (d) the ZXZ chain of equation~\eqref{eq:ZXZ} with $\Gamma = \Gamma_2 = 0.05$, $\lambda_1= \lambda_2=1$, and (e) the dimerized ZXZ chain with $\lambda_2= 0.6$, for system size $L =14$.
}
\label{fig:intro}
\end{figure*}

In particular, we show that the ZXZ model \cite{chen2014symmetry}, defined on an open one-dimensional chain with $L=2M$ sites as
\begin{align}
  H_{\textrm{SPT}} &= \lambda_1 \sum_{j=1}^{M-2} \sigma^z_{2j} \sigma^x_{2j+1}\sigma^z_{2j+2} +\lambda_2 \sum_{j=1}^{M-3} \sigma^z_{2j+1}\sigma^x_{2j+2}\sigma^z_{2j+3} \nonumber \\ &+\Gamma \sum_{j=1}^{L} \sigma^x_j + \Gamma_2 \sum_{j=1}^{L-1} \sigma^x_j \sigma^x_{j+1}, \label{eq:ZXZ}
\end{align}
can support a coherent edge qubit at any temperature, as long as it is dimerized with $\lambda_1 \neq \lambda_2$. 
More generally, the presence of this qubit owes to the existence of two long-lived, \emph{conjugate} boundary modes, the usual example of which are: $\{ \sigma^z_{\textrm{edge}}, \sigma^x_{\textrm{edge}} \}$ (Figure~\ref{fig:intro}).
Crucially, the dimerization breaks the $\mathbb{Z}_2$ swap symmetry between even and odd spins, but keeps the $\mathbb{Z}_2 \times \mathbb{Z}_2$ symmetry of the SPT phase intact. 
Finally, we propose an experimental realization of the model in a 1D Rydberg tweezer array and describe how the coherence of the edge qubit can be directly probed.

\emph{Edge decoupling of a classical bit---}
Before discussing the emergence of an edge qubit, we begin with some intuition for the edge decoupling mechanism.
This mechanism can already be illustrated for the \emph{classical} edge polarization of a quantum transverse-field Ising chain.
In particular, consider the Hamiltonian:
\begin{equation}
\label{eq:IsingHam}
H_{\textrm{Ising}}=-J\sum_{j=1}^{L-1} \sigma^z_j  \sigma^z_{j+1} - \Gamma \sum_{j=1}^{L-1} \sigma^x_j- J_2\sum_{j=1}^{L-2}\sigma^z_j\sigma^z_{j+2},\ 
\end{equation}
where $\sigma^{x/ z}$  are Pauli operators. It is well-known that the ground state of this system (for small enough $\Gamma$) is ferromagnetic. Dynamically, this is captured by the autocorrelation of the bulk magnetization,   $\langle \sigma^z(t) \sigma^z(0) \rangle_{T=0} \xrightarrow[t\rightarrow \infty]{\text{}} M^2 \neq 0$.

\begin{figure}
\hspace{-6mm} \includegraphics[width=3.6in]{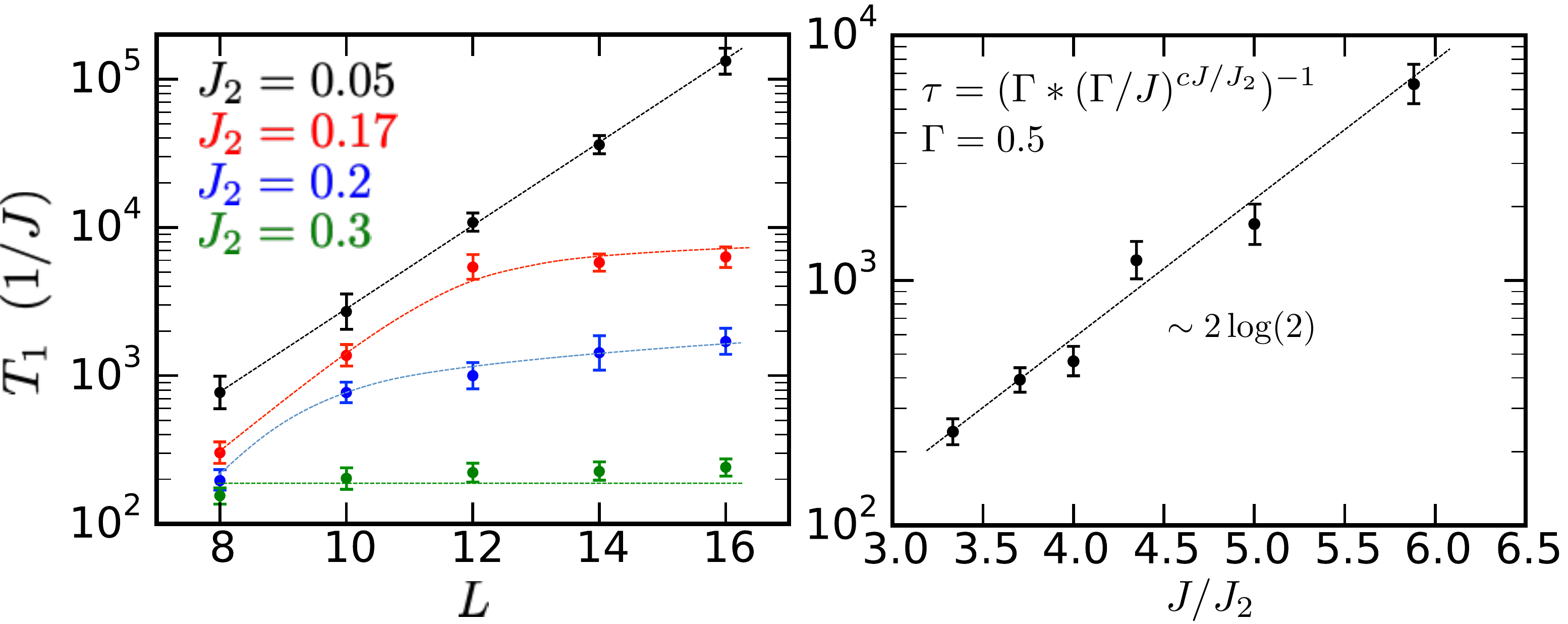}
\caption{%
The depolarization time $T_1$ of the classical edge bit of the Ising chain Eq.~\eqref{eq:IsingHam}.
(a) In finite size chains, $T_1$ is limited by perturbative processes which flip the spin at opposite ends of the chain. Thus, $T_1$ increases exponentially with $L$ until it saturates to its infinite volume limit.
(b) The saturated $T_1$ follows the exponential form $e^{c J/ J_2}$ as predicted by Eq.~\eqref{eq:decaytime}.
}
\label{fig:ising}
\end{figure}

At non-zero temperature,  quantum dynamics cause bulk observables to thermalize and such long-range order is lost. In particular, $\langle \sigma^z(t) \sigma^z(0) \rangle_T$ decays as $\sim e^{- t / \tau_{\textrm{bulk}}}$ with a timescale, $\tau_{\textrm{bulk}} \sim 1/\Gamma$, due to the propagation of bulk domain walls. 
Surprisingly, even at \emph{infinite} temperature in this interacting model, the edge magnetization, $\sigma^z_1$ (and, analogously, $\sigma^z_L$), can decay significantly more slowly. 
Indeed, for  $\Gamma, J_2 \ll J$, $\langle \sigma^z_1(t) \sigma^z_1(0) \rangle \sim e^{-t / \tau_{\textrm{edge}}}$ with 
\begin{equation}
  \frac{1}{\tau_{\textrm{edge}}} \sim \Gamma \left ( \frac{\Gamma}{ J} \right)^{c J / J_2}
  \label{eq:decaytime}
\end{equation}
as can be seen over two orders of magnitude in Figure~\ref{fig:ising}.

To understand the enhanced stability of the edge magnetization, one should consider the excited states of the Ising chain in terms of domain walls in the $\sigma^z$ configuration. 
For $J_2 = \Gamma = 0$, each domain wall costs energy $2J$. If the edge spin $\sigma_1^z$ flips, it changes the number of domain walls by $\pm 1$ and the energy by $\pm 2J$. In the bulk, turning on a perturbative transverse field $\Gamma$ can only change the number of domain walls by $0,\pm 2$. Thus, all finite-order perturbative processes (in $\Gamma$) which depolarize $\sigma_1^z$ are off-resonant by at least $\Delta E = \pm 2J$ and the edge magnetization cannot decay ($\tau_{\textrm{edge}} \to \infty$).

At finite $J_2$, the domain walls interact: any pair of domain walls gains (diagonal) energy $2J_2$ when they are neighbors.  Thus, it is possible to compensate the energy $\sim 2J$ of an extra domain wall by rearranging of order $n \sim J/J_2$ domain walls to sit next to one another.  Using this as the leading order on-shell process produces the exponential prethermal timescale of Eqn.~\eqref{eq:decaytime}.

The edge magnetization $\sigma_1^z$ in the Ising model thus constitutes a long-lived \emph{classical} bit at the boundary -- it resists depolarization from bulk dynamics even at high temperature. 
However, it is not a long-lived \emph{quantum} bit, which would also resist ``dephasing''. 
More precisely, any local operator conjugate to $\sigma^z_1$ (e.g.~$\sigma^x_1$ or $\sigma^y_1$) creates domain walls whose propagation leads to decay on a time scale $O(1/\Gamma)$, as illustrated in Figure~\ref{fig:intro}(c).




\emph{Edge decoupling of a quantum bit---}
%
%
%
Having developed intuition for the long-lived classical polarization, we turn to the edge qubit in the  $\mathbb{Z}_2 \times \mathbb{Z}_2$ SPT phase~\cite{Affleck87}.
For this qubit to remain coherent at high temperatures, the \emph{pair} of conjugate boundary modes corresponding to $\sigma^z$ and $\sigma^x$ must be long-lived. 
Naively, the simplest way to achieve this would be to generalize the ``domain-wall absorption arguments'' to each individual conjugate edge mode of the $\mathbb{Z}_2 \times \mathbb{Z}_2$ SPT. 
Unfortunately, there is an immediate complication: in the transverse-field Ising model there is only a single type of domain wall, whereas in a $\mathbb{Z}_2 \times \mathbb{Z}_2$ SPT, there are multiple types of excitations, leading to many more channels for depolarization and dephasing. 

The excitation structure of the $\mathbb{Z}_2 \times \mathbb{Z}_2$ SPT is most easily understood under duality \cite{suppinfo,Kennedy:92}. 
The ZXZ Hamiltonian Eq.~\eqref{eq:ZXZ} is dual to two coupled transverse-field Ising chains (one on the odd sites and the other on the even sites):
$
 H^\prime_{\textrm{SPT}} =  \lambda_1 \sum_{j=1}^{M-2}  \sigma^z_{2j} \sigma^z_{2j+2} + \Gamma \sum_{j=1}^{M}  \sigma^x_{2j}  + \lambda_2 \sum_{j=1}^{M-3}  \sigma^z_{2j+1} \sigma^z_{2j+3} + \Gamma \sum_{j=1}^{M-1}  \sigma^x_{2j+1}   + \Gamma_2 \sum_{i=1}^{L-1} \sigma^x_i \sigma^x_{i+1} \label{eq:ZXZising}
$.
When $\lambda_i$ are the dominant couplings, the SPT phase transforms to the global $\mathbb{Z}_2 \times \mathbb{Z}_2$ \emph{symmetry-broken} phase of the coupled Ising chains. 
It is clear from this dual picture that there is an additional $\mathbb{Z}_2$ ``swap'' symmetry when $\lambda_1 = \lambda_2$, which arises from exchanging the two Ising chains.
In addition, excitations of the original SPT correspond to different types of bulk domain walls in the dual symmetry-broken model. 

When the two chains are decoupled ($\Gamma_2 = 0$), their respective edge spins, $\sigma^z_1$ and $\sigma^z_2$, are protected from depolarization; however, these operators are not mutually conjugate. 
But the operator $\sigma^z_2 \prod_{i=1}^M \sigma^x_{2i-1}$ \emph{is} conjugate to $\sigma^z_1$, and it is long-lived, because the product over $\sigma^x_{2i-1}$ is simply the global spin-flip symmetry $G_\textrm{o}$ on odd sites. 
Under duality, the corresponding long-lived conjugate operators in the SPT are localized to the edge and  given by:
\begin{align}
  &\textrm{ZXZ} \hspace{70pt} \textrm{Ising} \times \textrm{Ising} \nonumber\\
\Sigma^x &= \sigma^x_1 \sigma^z_2 \qquad \longleftrightarrow \qquad \sigma^z_2 G_\textrm{o} \nonumber\\ 
\Sigma^y &= \sigma^y_1 \sigma^z_2 \qquad  \longleftrightarrow \qquad  i \sigma^z_1 \sigma^z_2 G_\textrm{o}  \nonumber\\                       
\Sigma^z &= \sigma^z_1 \hphantom{\sigma^z_1} \qquad \longleftrightarrow \qquad  \sigma^z_1. \label{eq:SPTedge} \end{align}

For decoupled transverse-field Ising chains, the depolarization of the edge requires the emission or absorption of a domain wall, which is an off-resonant process. 
However, if the chains are coupled, the interaction $\Gamma_2$ between them can depolarize the edge spins by transforming one type of domain wall into the other. 
In particular, if $\lambda_1 = \lambda_2$ then there are different types of domain walls with the same energy and the edge spin can immediately relax via on-shell domain-wall conversion. 
A similar physical argument can be used to explain the lack of a long-lived edge mode in the Potts model~\cite{Jermyn:2014, Moran17}. 

As an explicit example of this domain-wall conversion process, consider the following two configurations of spins at the edge of the coupled Ising chains:
\begin{align*}
  \uparrow \,\downarrow \,\downarrow\,\downarrow\,  \cdots &\qquad   \mathrel{\raisebox{-6 pt}{$\xRightarrow{\raisebox{4 pt}[0pt][0pt]{$\Gamma_2 \sigma^x_1 \sigma^x_2$}}$}}&  \downarrow \,\downarrow \,\downarrow\,\downarrow\,  \cdots \\
   \downarrow \,\downarrow \,\downarrow\,\downarrow\,  \cdots &\qquad &  \uparrow \,\downarrow \,\downarrow\,\downarrow\, \cdots 
\end{align*} 
Here, the upper (lower) row depicts the odd (even) spins.
On the left, there is a single broken Ising bond on the upper chain near the edge. 
The $\Gamma_2$ term can hop the broken bond from the upper chain to the lower flipping both edge spins. 
When $\lambda_1 = \lambda_2$, there is \emph{no difference} in energy between these two configurations, due to the $\mathbb{Z}_2$ swap symmetry. 
Thus, the protection of the edge spin fails at leading order in perturbation theory, and it quickly depolarizes [Fig.~\ref{fig:intro}(d)].
 
This suggests that a natural way to restore the edge spin lifetime is to dimerize the SPT with $\lambda_1 \neq \lambda_2$, which prevents the direct resonant conversion of one type of domain wall to the other.  
Consequently, the autocorrelation times of all the conjugate edge mode operators, $\Sigma^\alpha$, are exponentially long at infinite temperature, as can be seen in Figure~\ref{fig:intro}(e), in stark contrast with the transverse-field Ising chain, where only $\sigma^z_1$ has a long autocorrelation time [Fig.~\ref{fig:intro}(c)]. 
In the language of quantum information, this means that both depolarization \emph{and} dephasing are strongly suppressed and the edge mode constitutes a coherent qubit.

\emph{Interpretation via strong zero modes---}
Let us now turn to a deeper analytic understanding of the long-lived edge qubit.
We will demonstrate that it arises from the presence of two \emph{almost} strong edge zero modes, which exhibit significant overlap with $\Sigma^\alpha$.
Before diving in, we briefly summarize the properties of exact and almost strong zero modes~\cite{Fendley16, Kemp17, Else17, Vasiloiu18}.
An exact strong edge zero mode (SZM) is an operator which is  localized at the edge of the system, and commutes with the Hamiltonian up to a term exponentially small in system size.
A familiar example is the Majorana zero mode at the edge of the Kitaev chain~\cite{Kitaev01, Fendley12}. 
%
In principle, such SZMs can be constructed order by order in  perturbation theory. 
%
%
For systems with an exact SZM, the perturbative construction converges exponentially and need only be cut off by the finite size of the chain. 
In systems with an almost SZM, the same construction produces an asymptotic series which must be cut-off at some finite-order, beyond which the magnitude of the commutator with $H$ increases.
This cut-off produces the observed lifetimes in, e.g. Fig.~2(a).

\begin{figure}
 \includegraphics[width=0.85\columnwidth]{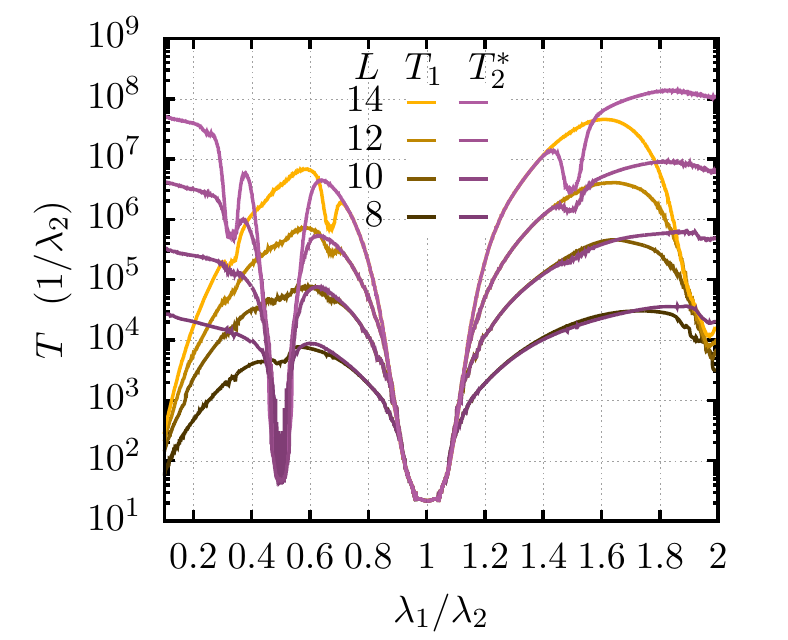}
\caption{The decay times $T_1$ and $T_2^*$ of the autocorrelators of the conjugate edge operators $\Sigma^z$ and $\Sigma^x$ respectively for the ZXZ model, Eq.~\eqref{eq:ZXZ}.
These results are calculated using exact diagonalisation at infinite temperature with $\Gamma = \Gamma_2 = 0.05$. 
The lifetimes increase exponentially with system size until a nearby resonance at rational $\lambda_1/\lambda_2$ causes them to saturate. The resonance at $\lambda_1 = \lambda_2$ is first order and thus there is no enhancement in the lifetime compared to bulk operators.}
\label{fig:zxz}
\end{figure}

Returning to the ZXZ model, we attempt to construct two conjugate SZMs by double expansion in $\Gamma$ and $\Gamma_2$, starting from the zeroth order terms: $\Psi_z^{(0)} = \Sigma^z$ and $\Psi_x^{(0)} = \Sigma^x$. 
We have explicitly constructed these up to fourth order and emphasize that the existence of two such \emph{conjugate} SZMs is highly non-trivial and does not occur, for example, in the transverse-field Ising model. 
Explicitly, the first order terms are
\begin{alignat*}{2}
  \Psi_z^{(1)} &=&& \frac{\Gamma}{\lambda_1} \sigma^x_{1} \sigma^x_{2} \sigma^z_{3}+\frac{ \Gamma_2}{\lambda_1^{2} - \lambda_{2}^{2}} (\lambda_1\sigma^x_{1} \sigma^z_{3}+ \lambda_{2} \sigma^y_{1} \sigma^y_{2} \sigma^x_{3} \sigma^z_{4}) \\
  \Psi_x^{(1)} &=&& \frac{\Gamma}{\lambda_2} \sigma^x_{1} \sigma^x_{2} \sigma^x_{3} \sigma^z_{4} - \frac{\Gamma_2}{\lambda_1^{2} - \lambda_2^{2}}(\lambda_2 \sigma^x_{2} \sigma^x_{3} \sigma^z_{4} + \lambda_1  \sigma^z_{1} \sigma^z_{2} \sigma^z_{3}) \\
  & && +\frac{\Gamma_2 \lambda_1}{4 \lambda_1^{2} - \lambda_2^{2}}\left(\sigma^y_{1} \sigma^z_{2} \sigma^y_{3} + (2\frac{\lambda_1}{\lambda_2}-\frac{\lambda_2}{\lambda_1}) \sigma^x_{1} \sigma^x_{2} \sigma^z_{4}  \right.\\ & && \hspace{2.2 cm}- \sigma^x_{1} \sigma^y_{2} \sigma^y_{3} \sigma^x_{4} \sigma^z_{5}   \left.- 2 \frac{\lambda_1}{\lambda_2} \sigma^y_{1} \sigma^y_{4} \sigma^z_{5}\right).
\end{alignat*}
While this expression may seem opaque, it contains much physics. 
For example, it already provides insight into the need for dimerization: 
for $|\lambda_1| = |\lambda_2|$, the expansion breaks down already at first order, since $\Psi^{(1)}$ diverges. 
This leads to a simple prediction --- the autocorrelation times of $\Sigma^z$ and $\Sigma^x$ should exhibit a dramatic reduction [compared to Fig.~\ref{fig:intro}(e)] when $|\lambda_1|/|\lambda_2|  = 1$.
This is indeed born out by our numerics, see Fig.~\ref{fig:zxz}.

More generally, poles in the SZM expansion correspond to physical resonances where the lifetimes becomes short. These can correspond to complicated physical processes.
For example, there is an additional pole in $\Psi_x^{(1)}$ at $2 |\lambda_1| = |\lambda_2|$ but not in $\Psi_z^{(1)}$. 
This corresponds to the large dip in the coherence time of $\Sigma^x$ around $\lambda_1 = \lambda_2 / 2$ in Figure~\ref{fig:zxz}.
Physically, a broken bond on the edge of the lower chain  can hop into the bulk of the upper chain, creating \emph{two} broken bonds. This process does not change the energy if $\lambda_1 = \lambda_2/2$. 
However, if a broken bond is on the edge of the upper chain, the $\Gamma_2$ coupling can only move it to the edge of the lower chain, so there is no corresponding dip in the coherence time of $\Sigma^z$!

At second order in the SZM expansion, further poles appear at $|\lambda_1|/|\lambda_2| = \sfrac{1}{3},2$ for $\Sigma^z$, and at $\sfrac{1}{3},\sfrac{3}{2}$ for $\Sigma^x$, the effects of which are directly visible in Fig.~\ref{fig:zxz}. 
The dip at \sfrac{2}{3} is due to a third order pole.  
In general, suppose $|\lambda_1|/|\lambda_2|  = p/q$ for integers, $p$ and $q$, without common prime factors. 
There is a resonance if the change in energy from flipping an edge spin under the diagonal part of the Hamiltonian ($\Gamma=\Gamma_2=0$) can be matched by flipping bulk spins. 
In the case of the coupled Ising chains, the dangerous processes involve flipping either of the edge spins individually or both together.
Let us factor the change in energy due to flipping a spin at site $j$ as $2 \lambda_1 \Delta_j/p$. Then at the edge, $\Delta_1 = \pm p$ and $\Delta_2 = \pm q$. 
In the bulk, the sum of $\Delta_j$ for $j > 2$ is an arbitrary linear combination of $2p$ and $2q$, which has even parity. 
Thus when $p$ is even, the edge spin flip cost $\Delta_1$ can be cancelled by the sum of bulk flips $\Delta_j$, and $\Sigma^z$ has a resonance. 
On the other hand, $p$ even forces $q$ odd, so $\Sigma^x$ does not have a resonance. 
When $q$ is even, the reverse is true. 
If instead both $p$ and $q$ are odd, the cost of flipping both edge spins $\Delta_1 + \Delta_2$ is even, so both edge operators suffer from a resonance at the same order.

The above analysis shows that there are no divergence-free, rational $|\lambda_1|/|\lambda_2|$ for both conjugate edge operators. 
Nevertheless, for large $p$ and $q$, the resonances only occur at high orders in perturbation theory, and the coherence time is significantly enhanced, as emphasized in Figure~\ref{fig:zxz}.
One might be tempted to bypass resonances entirely by choosing incommensurate $\lambda_1$ and $\lambda_2$, where there are formal results on bulk prethermal behaviour~\cite{Else19}.
However, the poles due to nearby resonances always produce large coefficients in the SZM expansion at sufficiently high order. 
Physically, the linewidth of the high-order resonances, clear in Fig.~\ref{fig:zxz}, may be interpreted in terms of the energy uncertainty of the domain walls involved in the processes that flip the edge spin~\cite{Kemp17}.

The SZM interpretation also naturally explains the height of the prethermal plateaus observed in the time-evolution of $\langle \Sigma^\alpha(t) \Sigma^\alpha(0) \rangle$ shown in Fig.~\ref{fig:intro}.
These are given by the appropriate overlap between the physical operators $\Sigma^\alpha$ and the SZM $\Psi^\alpha$. 
The decay from $\Sigma$ to $\Psi$ takes place on on a time-scale $t_{\rm plat}$ orders of magnitude less than the ultimate decay to zero. 
At infinite temperature, one can exactly calculate the overlap between $\Sigma^\alpha$ and $\Psi^\alpha$ as a normalized trace inner product: 
\begin{equation}
\langle \Sigma^\alpha(t_{\rm plat}) \Sigma^\alpha(0) \rangle\big\rvert_{T=\infty} =\frac{1}{2^{L}}\Tr(\Sigma^\alpha \Psi^\alpha)^2.\label{eq:traceovelap}
\end{equation}.


\emph{Experimental realization.---}%
A particularly direct experimental realization of our proposal can be implemented in a 1D optical tweezer array (Fig.~\ref{fig:intro}) of single alkali or alkaline-earth atoms \cite{barredo2016atom,endres2016atom,norcia2018microscopic,cooper2018alkaline,saskin2019narrow}. 
Such systems have recently emerged as powerful platforms for building up many-body quantum systems atom-by-atom. 
Here, we envision the effective spin degree of freedom in Eq.~\eqref{eq:ZXZ} to be formed by two hyperfine atomic ground states.
The most natural Hamiltonian available in such a system is a long-range transverse field Ising model: $H = \sum_{i=1}^{N} h_{i} \sigma_{i}^{x} + \sum_{i=1}^{N-1}  \lambda_{i} \sigma_{i}^{z}\sigma_{i+1}^{z}$.
The Ising interaction can be generated by dressing the ground hyperfine state with an excited Rydberg state using a far-detuned laser~\cite{vanBijnen15,Glaetzle15,Goldschmidt16,Zeiher16,zeiher2017coherent}; the resulting Rydberg blockade induces strong effective spin-spin interactions with a range on the order of a few microns \cite{blockade}. 
The transverse field can be implemented by resonant Raman coupling. 
Finally, by using techniques from Floquet engineering, it is possible to realize the dimerized ZXZ Hamiltonian stroboscopically \cite{Iadecola15, Potirniche17}.
In particular, by periodically modulating the Ising coupling as $ \omega \cos(\omega t)\lambda_{i} \sigma_{i}^{z}\sigma_{i+1}^{z}$ \cite{drivingf}, one generates (at leading order in a Floquet-Magnus expansion) dynamics that are governed by an effective Floquet Hamiltonian \cite{Potirniche17}:
\begin{align}\label{eq:stroboscopicHamiltonian}
H_{\mathrm{F}} &=& \sum_{i=1}^{N} h_{i}a(\lambda_{1}, \lambda_{2}) \sigma_{i}^{x} - \sum_{i=2}^{N-1}h_{i}b(\lambda_{1}, \lambda_{2}) \sigma_{i-1}^{z}\sigma_{i}^{x}\sigma_{i+1}^{z} ,
\end{align}
where $a(\lambda_{1}, \lambda_{2}) = \frac{1}{2} \left[J_{0}\left(2(\lambda_{1} - \lambda_{2})\right) + J_{0}(2(\lambda_{1}+\lambda_{2})) \right]$,  $b(\lambda_{1},\lambda_{2})  = J_{0}(2(\lambda_{1}-\lambda_{2})) - a(\lambda_{1}, \lambda_{2})$ and $J_{0}(x)$ is a Bessel function of the first kind \cite{Potirniche17,floquetinteractions}.
Crucially, dimerization of the transverse field $h_i$ is inherited by the ZXZ coupling term in the Floquet Hamiltonian.

The relatively large inter-atom spacing of a typical Rydberg tweezer array naturally enables experiments to probe the long lifetime associated with our proposed coherent edge qubit. 
As a concrete protocol, 1) initialize the Rydberg spin chain in a randomly oriented product state (i.e.~effectively at infinite temperature); 
2) Projectively initialize the edge qubit by measuring $\Sigma^z$; 
3) Allow the system to evolve time for a time $t$ and 4) measure $\Sigma^z$ again to obtain the correlation function $\braket{\Sigma^z(t) \Sigma^z(0)}$ by averaging over multiple runs. 
%
%
The decay of the $\braket{\Sigma^z(t) \Sigma^z(0)}$ correlator measures the lifetime of $z$-polarization.
In order to demonstrate a coherent quantum bit, one must perform the same exact procedure for the conjugate $\Sigma^x$ operator. 
In the language of atomic spectroscopy, this is analogous to performing a Ramsey sequence to probe the dephasing time, $T_2^*$, of the qubit (as opposed to the $\Sigma^z$ autocorrelator, which probes the depolarization time, $T_1$).

In order to ensure that one can indeed choose parameters for the Floquet Hamiltonian Eq.~\eqref{eq:stroboscopicHamiltonian} that  realize a dimerized ZXZ model with long edge coherence, we have numerically simulated the Floquet Hamiltonian for an $L=14$ spin chain \cite{suppinfo}. 
In addition to demonstrating that our proposal works for a broad range of parameters, our numerics indicate that the difference between edge and bulk spin autocorrelators can be distinguished on time-scales much shorter than the typical lifetime of the Rydberg-dressed state~\cite{Lee17}.
%
%

Our work opens the door to a number of intriguing future directions. 
First, by exploring other symmetry regimes, higher spatial dimensions, and models with unbounded local Hilbert spaces, it may be possible to extend our mechanism for edge mode stability to a more generic setting.
Second, while we have focused on an experimental proposal based on Floquet engineering, it would be interesting to investigate the prethermal dynamics of a coupled 1D Rydberg ladder; this geometry naturally exhibits the same symmetries as the ZXZ model and thus might provide a simpler route to realizing symmetry enhanced edge modes.
Finally, building on techniques originally developed in the context of many-body localized SPT phases~\cite{Yao15,choi2015quantum}, it would be interesting to explore hybrid quantum information protocols where symmetry enhanced edge qubits play the role of robust quantum memories.

\emph{Note added:} A similar model with $\mathbb{Z}_2 \times \mathbb{Z}_2$ SPT order has been considered by Parker \textit{et al.}~\cite{Parker19} However, they focus on the effects of the SPT phase on a single SZM in a nearby proximate symmetry-broken phase, rather than considering the conjugate edge modes in the SPT itself.

\emph{Acknowledgments---}The authors would like to thank Soonwon Choi, Paul Fendley, Francisco Machado, Johannes Zeiher, and Immanuel Bloch for stimulating discussions. 
C.R.L. acknowledges support from the NSF through grant PHY-1752727.
J.K. acknowledges support through the NSF QII-TAQS program (grant number 1936100).
N.Y.Y. acknowledges support from the ARO through the MURI program (grant W911NF-17-1-0323).
This work was performed in part at the Aspen Center for Physics, which is supported by National Science Foundation grant PHY-1607611, and at the KITP, supported by Grant No. NSF PHY-1748958.
Any opinion, findings, and conclusions or recommendations expressed in this material are those of the authors and do not necessarily reflect the views of the NSF.

\bibliography{szm}

\begin{thebibliography}{42}%
\makeatletter
\providecommand \@ifxundefined [1]{%
 \@ifx{#1\undefined}
}%
\providecommand \@ifnum [1]{%
 \ifnum #1\expandafter \@firstoftwo
 \else \expandafter \@secondoftwo
 \fi
}%
\providecommand \@ifx [1]{%
 \ifx #1\expandafter \@firstoftwo
 \else \expandafter \@secondoftwo
 \fi
}%
\providecommand \natexlab [1]{#1}%
\providecommand \enquote  [1]{``#1''}%
\providecommand \bibnamefont  [1]{#1}%
\providecommand \bibfnamefont [1]{#1}%
\providecommand \citenamefont [1]{#1}%
\providecommand \href@noop [0]{\@secondoftwo}%
\providecommand \href [0]{\begingroup \@sanitize@url \@href}%
\providecommand \@href[1]{\@@startlink{#1}\@@href}%
\providecommand \@@href[1]{\endgroup#1\@@endlink}%
\providecommand \@sanitize@url [0]{\catcode `\\12\catcode `\$12\catcode
  `\&12\catcode `\#12\catcode `\^12\catcode `\_12\catcode `\%12\relax}%
\providecommand \@@startlink[1]{}%
\providecommand \@@endlink[0]{}%
\providecommand \url  [0]{\begingroup\@sanitize@url \@url }%
\providecommand \@url [1]{\endgroup\@href {#1}{\urlprefix }}%
\providecommand \urlprefix  [0]{URL }%
\providecommand \Eprint [0]{\href }%
\providecommand \doibase [0]{http://dx.doi.org/}%
\providecommand \selectlanguage [0]{\@gobble}%
\providecommand \bibinfo  [0]{\@secondoftwo}%
\providecommand \bibfield  [0]{\@secondoftwo}%
\providecommand \translation [1]{[#1]}%
\providecommand \BibitemOpen [0]{}%
\providecommand \bibitemStop [0]{}%
\providecommand \bibitemNoStop [0]{.\EOS\space}%
\providecommand \EOS [0]{\spacefactor3000\relax}%
\providecommand \BibitemShut  [1]{\csname bibitem#1\endcsname}%
\let\auto@bib@innerbib\@empty
\bibitem [{\citenamefont {Nandkishore}\ and\ \citenamefont
  {Huse}(2015)}]{nandkishore2015many}%
  \BibitemOpen
  \bibfield  {author} {\bibinfo {author} {\bibfnamefont {R.}~\bibnamefont
  {Nandkishore}}\ and\ \bibinfo {author} {\bibfnamefont {D.~A.}\ \bibnamefont
  {Huse}},\ }\href@noop {} {\bibfield  {journal} {\bibinfo  {journal} {Annu.
  Rev. Condens. Matter Phys.}\ }\textbf {\bibinfo {volume} {6}},\ \bibinfo
  {pages} {15} (\bibinfo {year} {2015})}\BibitemShut {NoStop}%
\bibitem [{\citenamefont {Abanin}\ and\ \citenamefont
  {Papi{\'c}}(2017)}]{abanin2017recent}%
  \BibitemOpen
  \bibfield  {author} {\bibinfo {author} {\bibfnamefont {D.~A.}\ \bibnamefont
  {Abanin}}\ and\ \bibinfo {author} {\bibfnamefont {Z.}~\bibnamefont
  {Papi{\'c}}},\ }\href@noop {} {\bibfield  {journal} {\bibinfo  {journal}
  {Annalen der Physik}\ }\textbf {\bibinfo {volume} {529}},\ \bibinfo {pages}
  {1700169} (\bibinfo {year} {2017})}\BibitemShut {NoStop}%
\bibitem [{\citenamefont {Chandran}\ \emph {et~al.}(2014)\citenamefont
  {Chandran}, \citenamefont {Khemani}, \citenamefont {Laumann},\ and\
  \citenamefont {Sondhi}}]{Chandran14}%
  \BibitemOpen
  \bibfield  {author} {\bibinfo {author} {\bibfnamefont {A.}~\bibnamefont
  {Chandran}}, \bibinfo {author} {\bibfnamefont {V.}~\bibnamefont {Khemani}},
  \bibinfo {author} {\bibfnamefont {C.~R.}\ \bibnamefont {Laumann}}, \ and\
  \bibinfo {author} {\bibfnamefont {S.~L.}\ \bibnamefont {Sondhi}},\ }\href
  {\doibase 10.1103/PhysRevB.89.144201} {\bibfield  {journal} {\bibinfo
  {journal} {Phys. Rev. B}\ }\textbf {\bibinfo {volume} {89}},\ \bibinfo
  {pages} {144201} (\bibinfo {year} {2014})}\BibitemShut {NoStop}%
\bibitem [{\citenamefont {Potter}\ and\ \citenamefont
  {Vishwanath}(2015)}]{Potter15}%
  \BibitemOpen
  \bibfield  {author} {\bibinfo {author} {\bibfnamefont {A.~C.}\ \bibnamefont
  {Potter}}\ and\ \bibinfo {author} {\bibfnamefont {A.}~\bibnamefont
  {Vishwanath}},\ }\href@noop {} {\  (\bibinfo {year} {2015})},\ \Eprint
  {http://arxiv.org/abs/1506.00592} {arXiv:1506.00592 [cond-mat.dis-nn]}
  \BibitemShut {NoStop}%
\bibitem [{\citenamefont {Bahri}\ \emph {et~al.}(2015)\citenamefont {Bahri},
  \citenamefont {Vosk}, \citenamefont {Altman},\ and\ \citenamefont
  {Vishwanath}}]{Bahri15}%
  \BibitemOpen
  \bibfield  {author} {\bibinfo {author} {\bibfnamefont {Y.}~\bibnamefont
  {Bahri}}, \bibinfo {author} {\bibfnamefont {R.}~\bibnamefont {Vosk}},
  \bibinfo {author} {\bibfnamefont {E.}~\bibnamefont {Altman}}, \ and\ \bibinfo
  {author} {\bibfnamefont {A.}~\bibnamefont {Vishwanath}},\ }\href
  {http://dx.doi.org/10.1038/ncomms8341} {\bibfield  {journal} {\bibinfo
  {journal} {Nature Communications}\ }\textbf {\bibinfo {volume} {6}},\
  \bibinfo {pages} {7341 EP } (\bibinfo {year} {2015})},\ \Eprint
  {http://arxiv.org/abs/1307.4092} {1307.4092} \BibitemShut {NoStop}%
\bibitem [{\citenamefont {Yao}\ \emph {et~al.}(2015)\citenamefont {Yao},
  \citenamefont {Laumann},\ and\ \citenamefont {Vishwanath}}]{Yao15}%
  \BibitemOpen
  \bibfield  {author} {\bibinfo {author} {\bibfnamefont {N.~Y.}\ \bibnamefont
  {Yao}}, \bibinfo {author} {\bibfnamefont {C.~R.}\ \bibnamefont {Laumann}}, \
  and\ \bibinfo {author} {\bibfnamefont {A.}~\bibnamefont {Vishwanath}},\
  }\href@noop {} {\bibfield  {journal} {\bibinfo  {journal} {arXiv:1508.06995}\
  } (\bibinfo {year} {2015})}\BibitemShut {NoStop}%
\bibitem [{\citenamefont {Schreiber}\ \emph {et~al.}(2015)\citenamefont
  {Schreiber}, \citenamefont {Hodgman}, \citenamefont {Bordia}, \citenamefont
  {L{\"u}schen}, \citenamefont {Fischer}, \citenamefont {Vosk}, \citenamefont
  {Altman}, \citenamefont {Schneider},\ and\ \citenamefont
  {Bloch}}]{schreiber2015observation}%
  \BibitemOpen
  \bibfield  {author} {\bibinfo {author} {\bibfnamefont {M.}~\bibnamefont
  {Schreiber}}, \bibinfo {author} {\bibfnamefont {S.~S.}\ \bibnamefont
  {Hodgman}}, \bibinfo {author} {\bibfnamefont {P.}~\bibnamefont {Bordia}},
  \bibinfo {author} {\bibfnamefont {H.~P.}\ \bibnamefont {L{\"u}schen}},
  \bibinfo {author} {\bibfnamefont {M.~H.}\ \bibnamefont {Fischer}}, \bibinfo
  {author} {\bibfnamefont {R.}~\bibnamefont {Vosk}}, \bibinfo {author}
  {\bibfnamefont {E.}~\bibnamefont {Altman}}, \bibinfo {author} {\bibfnamefont
  {U.}~\bibnamefont {Schneider}}, \ and\ \bibinfo {author} {\bibfnamefont
  {I.}~\bibnamefont {Bloch}},\ }\href@noop {} {\bibfield  {journal} {\bibinfo
  {journal} {Science}\ }\textbf {\bibinfo {volume} {349}},\ \bibinfo {pages}
  {842} (\bibinfo {year} {2015})}\BibitemShut {NoStop}%
\bibitem [{\citenamefont {Choi}\ \emph {et~al.}(2016)\citenamefont {Choi},
  \citenamefont {Hild}, \citenamefont {Zeiher}, \citenamefont {Schau{\ss}},
  \citenamefont {Rubio-Abadal}, \citenamefont {Yefsah}, \citenamefont
  {Khemani}, \citenamefont {Huse}, \citenamefont {Bloch},\ and\ \citenamefont
  {Gross}}]{choi2016exploring}%
  \BibitemOpen
  \bibfield  {author} {\bibinfo {author} {\bibfnamefont {J.-y.}\ \bibnamefont
  {Choi}}, \bibinfo {author} {\bibfnamefont {S.}~\bibnamefont {Hild}}, \bibinfo
  {author} {\bibfnamefont {J.}~\bibnamefont {Zeiher}}, \bibinfo {author}
  {\bibfnamefont {P.}~\bibnamefont {Schau{\ss}}}, \bibinfo {author}
  {\bibfnamefont {A.}~\bibnamefont {Rubio-Abadal}}, \bibinfo {author}
  {\bibfnamefont {T.}~\bibnamefont {Yefsah}}, \bibinfo {author} {\bibfnamefont
  {V.}~\bibnamefont {Khemani}}, \bibinfo {author} {\bibfnamefont {D.~A.}\
  \bibnamefont {Huse}}, \bibinfo {author} {\bibfnamefont {I.}~\bibnamefont
  {Bloch}}, \ and\ \bibinfo {author} {\bibfnamefont {C.}~\bibnamefont
  {Gross}},\ }\href@noop {} {\bibfield  {journal} {\bibinfo  {journal}
  {Science}\ }\textbf {\bibinfo {volume} {352}},\ \bibinfo {pages} {1547}
  (\bibinfo {year} {2016})}\BibitemShut {NoStop}%
\bibitem [{\citenamefont {Smith}\ \emph {et~al.}(2016)\citenamefont {Smith},
  \citenamefont {Lee}, \citenamefont {Richerme}, \citenamefont {Neyenhuis},
  \citenamefont {Hess}, \citenamefont {Hauke}, \citenamefont {Heyl},
  \citenamefont {Huse},\ and\ \citenamefont {Monroe}}]{smith2016many}%
  \BibitemOpen
  \bibfield  {author} {\bibinfo {author} {\bibfnamefont {J.}~\bibnamefont
  {Smith}}, \bibinfo {author} {\bibfnamefont {A.}~\bibnamefont {Lee}}, \bibinfo
  {author} {\bibfnamefont {P.}~\bibnamefont {Richerme}}, \bibinfo {author}
  {\bibfnamefont {B.}~\bibnamefont {Neyenhuis}}, \bibinfo {author}
  {\bibfnamefont {P.~W.}\ \bibnamefont {Hess}}, \bibinfo {author}
  {\bibfnamefont {P.}~\bibnamefont {Hauke}}, \bibinfo {author} {\bibfnamefont
  {M.}~\bibnamefont {Heyl}}, \bibinfo {author} {\bibfnamefont {D.~A.}\
  \bibnamefont {Huse}}, \ and\ \bibinfo {author} {\bibfnamefont
  {C.}~\bibnamefont {Monroe}},\ }\href@noop {} {\bibfield  {journal} {\bibinfo
  {journal} {Nature Physics}\ }\textbf {\bibinfo {volume} {12}},\ \bibinfo
  {pages} {907} (\bibinfo {year} {2016})}\BibitemShut {NoStop}%
\bibitem [{\citenamefont {L{\"u}schen}\ \emph {et~al.}(2017)\citenamefont
  {L{\"u}schen}, \citenamefont {Bordia}, \citenamefont {Hodgman}, \citenamefont
  {Schreiber}, \citenamefont {Sarkar}, \citenamefont {Daley}, \citenamefont
  {Fischer}, \citenamefont {Altman}, \citenamefont {Bloch},\ and\ \citenamefont
  {Schneider}}]{luschen2017signatures}%
  \BibitemOpen
  \bibfield  {author} {\bibinfo {author} {\bibfnamefont {H.~P.}\ \bibnamefont
  {L{\"u}schen}}, \bibinfo {author} {\bibfnamefont {P.}~\bibnamefont {Bordia}},
  \bibinfo {author} {\bibfnamefont {S.~S.}\ \bibnamefont {Hodgman}}, \bibinfo
  {author} {\bibfnamefont {M.}~\bibnamefont {Schreiber}}, \bibinfo {author}
  {\bibfnamefont {S.}~\bibnamefont {Sarkar}}, \bibinfo {author} {\bibfnamefont
  {A.~J.}\ \bibnamefont {Daley}}, \bibinfo {author} {\bibfnamefont {M.~H.}\
  \bibnamefont {Fischer}}, \bibinfo {author} {\bibfnamefont {E.}~\bibnamefont
  {Altman}}, \bibinfo {author} {\bibfnamefont {I.}~\bibnamefont {Bloch}}, \
  and\ \bibinfo {author} {\bibfnamefont {U.}~\bibnamefont {Schneider}},\
  }\href@noop {} {\bibfield  {journal} {\bibinfo  {journal} {Physical Review
  X}\ }\textbf {\bibinfo {volume} {7}},\ \bibinfo {pages} {011034} (\bibinfo
  {year} {2017})}\BibitemShut {NoStop}%
\bibitem [{\citenamefont {Bordia}\ \emph {et~al.}(2017)\citenamefont {Bordia},
  \citenamefont {L{\"u}schen}, \citenamefont {Schneider}, \citenamefont
  {Knap},\ and\ \citenamefont {Bloch}}]{bordia2017periodically}%
  \BibitemOpen
  \bibfield  {author} {\bibinfo {author} {\bibfnamefont {P.}~\bibnamefont
  {Bordia}}, \bibinfo {author} {\bibfnamefont {H.}~\bibnamefont {L{\"u}schen}},
  \bibinfo {author} {\bibfnamefont {U.}~\bibnamefont {Schneider}}, \bibinfo
  {author} {\bibfnamefont {M.}~\bibnamefont {Knap}}, \ and\ \bibinfo {author}
  {\bibfnamefont {I.}~\bibnamefont {Bloch}},\ }\href@noop {} {\bibfield
  {journal} {\bibinfo  {journal} {Nature Physics}\ }\textbf {\bibinfo {volume}
  {13}},\ \bibinfo {pages} {460} (\bibinfo {year} {2017})}\BibitemShut
  {NoStop}%
\bibitem [{\citenamefont {Chen}\ \emph {et~al.}(2014)\citenamefont {Chen},
  \citenamefont {Lu},\ and\ \citenamefont {Vishwanath}}]{chen2014symmetry}%
  \BibitemOpen
  \bibfield  {author} {\bibinfo {author} {\bibfnamefont {X.}~\bibnamefont
  {Chen}}, \bibinfo {author} {\bibfnamefont {Y.-M.}\ \bibnamefont {Lu}}, \ and\
  \bibinfo {author} {\bibfnamefont {A.}~\bibnamefont {Vishwanath}},\
  }\href@noop {} {\bibfield  {journal} {\bibinfo  {journal} {Nature
  communications}\ }\textbf {\bibinfo {volume} {5}},\ \bibinfo {pages} {3507}
  (\bibinfo {year} {2014})}\BibitemShut {NoStop}%
\bibitem [{\citenamefont {Affleck}\ \emph {et~al.}(1987)\citenamefont
  {Affleck}, \citenamefont {Kennedy}, \citenamefont {Lieb},\ and\ \citenamefont
  {Tasaki}}]{Affleck87}%
  \BibitemOpen
  \bibfield  {author} {\bibinfo {author} {\bibfnamefont {I.}~\bibnamefont
  {Affleck}}, \bibinfo {author} {\bibfnamefont {T.}~\bibnamefont {Kennedy}},
  \bibinfo {author} {\bibfnamefont {E.~H.}\ \bibnamefont {Lieb}}, \ and\
  \bibinfo {author} {\bibfnamefont {H.}~\bibnamefont {Tasaki}},\ }\href
  {\doibase 10.1103/PhysRevLett.59.799} {\bibfield  {journal} {\bibinfo
  {journal} {Phys. Rev. Lett.}\ }\textbf {\bibinfo {volume} {59}},\ \bibinfo
  {pages} {799} (\bibinfo {year} {1987})}\BibitemShut {NoStop}%
\bibitem [{sup()}]{suppinfo}%
  \BibitemOpen
  \href@noop {} {}\bibinfo {note} {See Supplementary Material for additional
  details.}\BibitemShut {Stop}%
\bibitem [{\citenamefont {Kennedy}\ and\ \citenamefont
  {Tasaki}(1992)}]{Kennedy:92}%
  \BibitemOpen
  \bibfield  {author} {\bibinfo {author} {\bibfnamefont {T.}~\bibnamefont
  {Kennedy}}\ and\ \bibinfo {author} {\bibfnamefont {H.}~\bibnamefont
  {Tasaki}},\ }\href {\doibase 10.1103/PhysRevB.45.304} {\bibfield  {journal}
  {\bibinfo  {journal} {Phys. Rev. B}\ }\textbf {\bibinfo {volume} {45}},\
  \bibinfo {pages} {304} (\bibinfo {year} {1992})}\BibitemShut {NoStop}%
\bibitem [{\citenamefont {Jermyn}\ \emph {et~al.}(2014)\citenamefont {Jermyn},
  \citenamefont {Mong}, \citenamefont {Alicea},\ and\ \citenamefont
  {Fendley}}]{Jermyn:2014}%
  \BibitemOpen
  \bibfield  {author} {\bibinfo {author} {\bibfnamefont {A.}~\bibnamefont
  {Jermyn}}, \bibinfo {author} {\bibfnamefont {R.}~\bibnamefont {Mong}},
  \bibinfo {author} {\bibfnamefont {J.}~\bibnamefont {Alicea}}, \ and\ \bibinfo
  {author} {\bibfnamefont {P.}~\bibnamefont {Fendley}},\ }\href {\doibase
  10.1103/PhysRevB.90.165106} {\bibfield  {journal} {\bibinfo  {journal} {Phys.
  Rev. B}\ }\textbf {\bibinfo {volume} {90}},\ \bibinfo {pages} {165106}
  (\bibinfo {year} {2014})},\ \Eprint {http://arxiv.org/abs/arXiv:1407.6376}
  {arXiv:1407.6376} \BibitemShut {NoStop}%
\bibitem [{\citenamefont {Moran}\ \emph {et~al.}(2017)\citenamefont {Moran},
  \citenamefont {Pellegrino}, \citenamefont {Slingerland},\ and\ \citenamefont
  {Kells}}]{Moran17}%
  \BibitemOpen
  \bibfield  {author} {\bibinfo {author} {\bibfnamefont {N.}~\bibnamefont
  {Moran}}, \bibinfo {author} {\bibfnamefont {D.}~\bibnamefont {Pellegrino}},
  \bibinfo {author} {\bibfnamefont {J.~K.}\ \bibnamefont {Slingerland}}, \ and\
  \bibinfo {author} {\bibfnamefont {G.}~\bibnamefont {Kells}},\ }\href
  {\doibase 10.1103/PhysRevB.95.235127} {\bibfield  {journal} {\bibinfo
  {journal} {Phys. Rev. B}\ }\textbf {\bibinfo {volume} {95}},\ \bibinfo
  {pages} {235127} (\bibinfo {year} {2017})}\BibitemShut {NoStop}%
\bibitem [{\citenamefont {Fendley}(2016)}]{Fendley16}%
  \BibitemOpen
  \bibfield  {author} {\bibinfo {author} {\bibfnamefont {P.}~\bibnamefont
  {Fendley}},\ }\href {\doibase 10.1088/1751-8113/49/30/30LT01} {\bibfield
  {journal} {\bibinfo  {journal} {J. Phys.}\ }\textbf {\bibinfo {volume}
  {A49}},\ \bibinfo {pages} {30LT01} (\bibinfo {year} {2016})},\ \Eprint
  {http://arxiv.org/abs/1512.03441} {arXiv:1512.03441} \BibitemShut {NoStop}%
\bibitem [{\citenamefont {Kemp}\ \emph {et~al.}(2017)\citenamefont {Kemp},
  \citenamefont {Yao}, \citenamefont {Laumann},\ and\ \citenamefont
  {Fendley}}]{Kemp17}%
  \BibitemOpen
  \bibfield  {author} {\bibinfo {author} {\bibfnamefont {J.}~\bibnamefont
  {Kemp}}, \bibinfo {author} {\bibfnamefont {N.~Y.}\ \bibnamefont {Yao}},
  \bibinfo {author} {\bibfnamefont {C.~R.}\ \bibnamefont {Laumann}}, \ and\
  \bibinfo {author} {\bibfnamefont {P.}~\bibnamefont {Fendley}},\ }\href
  {\doibase 10.1088/1742-5468/aa73f0} {\bibfield  {journal} {\bibinfo
  {journal} {J. Stat. Mech.}\ }\textbf {\bibinfo {volume} {2017}},\ \bibinfo
  {pages} {063105} (\bibinfo {year} {2017})},\ \bibinfo {note} {arXiv:
  1701.00797}\BibitemShut {NoStop}%
\bibitem [{\citenamefont {Else}\ \emph {et~al.}(2017)\citenamefont {Else},
  \citenamefont {Fendley}, \citenamefont {Kemp},\ and\ \citenamefont
  {Nayak}}]{Else17}%
  \BibitemOpen
  \bibfield  {author} {\bibinfo {author} {\bibfnamefont {D.~V.}\ \bibnamefont
  {Else}}, \bibinfo {author} {\bibfnamefont {P.}~\bibnamefont {Fendley}},
  \bibinfo {author} {\bibfnamefont {J.}~\bibnamefont {Kemp}}, \ and\ \bibinfo
  {author} {\bibfnamefont {C.}~\bibnamefont {Nayak}},\ }\href {\doibase
  10.1103/PhysRevX.7.041062} {\bibfield  {journal} {\bibinfo  {journal} {Phys.
  Rev. X}\ }\textbf {\bibinfo {volume} {7}},\ \bibinfo {pages} {041062}
  (\bibinfo {year} {2017})}\BibitemShut {NoStop}%
\bibitem [{\citenamefont {Vasiloiu}\ \emph {et~al.}(2018)\citenamefont
  {Vasiloiu}, \citenamefont {Carollo},\ and\ \citenamefont
  {Garrahan}}]{Vasiloiu18}%
  \BibitemOpen
  \bibfield  {author} {\bibinfo {author} {\bibfnamefont {L.~M.}\ \bibnamefont
  {Vasiloiu}}, \bibinfo {author} {\bibfnamefont {F.}~\bibnamefont {Carollo}}, \
  and\ \bibinfo {author} {\bibfnamefont {J.~P.}\ \bibnamefont {Garrahan}},\
  }\href {\doibase 10.1103/PhysRevB.98.094308} {\bibfield  {journal} {\bibinfo
  {journal} {Phys. Rev. B}\ }\textbf {\bibinfo {volume} {98}},\ \bibinfo
  {pages} {094308} (\bibinfo {year} {2018})}\BibitemShut {NoStop}%
\bibitem [{\citenamefont {{Kitaev}}(2001)}]{Kitaev01}%
  \BibitemOpen
  \bibfield  {author} {\bibinfo {author} {\bibfnamefont {A.~Y.}\ \bibnamefont
  {{Kitaev}}},\ }\href {\doibase 10.1070/1063-7869/44/10S/S29} {\bibfield
  {journal} {\bibinfo  {journal} {Physics Uspekhi}\ }\textbf {\bibinfo {volume}
  {44}},\ \bibinfo {pages} {131} (\bibinfo {year} {2001})},\ \Eprint
  {http://arxiv.org/abs/arXiv:cond-mat/0010440} {arXiv:cond-mat/0010440}
  \BibitemShut {NoStop}%
\bibitem [{\citenamefont {{Fendley}}(2012)}]{Fendley12}%
  \BibitemOpen
  \bibfield  {author} {\bibinfo {author} {\bibfnamefont {P.}~\bibnamefont
  {{Fendley}}},\ }\href {\doibase 10.1088/1742-5468/2012/11/P11020} {\bibfield
  {journal} {\bibinfo  {journal} {J.~Stat.~Mech.}\ }\textbf {\bibinfo {volume}
  {11}},\ \bibinfo {pages} {20} (\bibinfo {year} {2012})},\ \Eprint
  {http://arxiv.org/abs/arXiv:1209.0472} {arXiv:1209.0472} \BibitemShut
  {NoStop}%
\bibitem [{\citenamefont {Else}\ \emph {et~al.}(2019)\citenamefont {Else},
  \citenamefont {Ho},\ and\ \citenamefont {Dumitrescu}}]{Else19}%
  \BibitemOpen
  \bibfield  {author} {\bibinfo {author} {\bibfnamefont {D.~V.}\ \bibnamefont
  {Else}}, \bibinfo {author} {\bibfnamefont {W.~W.}\ \bibnamefont {Ho}}, \ and\
  \bibinfo {author} {\bibfnamefont {P.~T.}\ \bibnamefont {Dumitrescu}},\
  }\href@noop {} {\bibfield  {journal} {\bibinfo  {journal} {arXiv preprint}\ }
  (\bibinfo {year} {2019})},\ \Eprint {http://arxiv.org/abs/1910.03584}
  {arXiv:1910.03584 [cond-mat.str-el]} \BibitemShut {NoStop}%
\bibitem [{\citenamefont {Barredo}\ \emph {et~al.}(2016)\citenamefont
  {Barredo}, \citenamefont {De~L{\'e}s{\'e}leuc}, \citenamefont {Lienhard},
  \citenamefont {Lahaye},\ and\ \citenamefont {Browaeys}}]{barredo2016atom}%
  \BibitemOpen
  \bibfield  {author} {\bibinfo {author} {\bibfnamefont {D.}~\bibnamefont
  {Barredo}}, \bibinfo {author} {\bibfnamefont {S.}~\bibnamefont
  {De~L{\'e}s{\'e}leuc}}, \bibinfo {author} {\bibfnamefont {V.}~\bibnamefont
  {Lienhard}}, \bibinfo {author} {\bibfnamefont {T.}~\bibnamefont {Lahaye}}, \
  and\ \bibinfo {author} {\bibfnamefont {A.}~\bibnamefont {Browaeys}},\
  }\href@noop {} {\bibfield  {journal} {\bibinfo  {journal} {Science}\ }\textbf
  {\bibinfo {volume} {354}},\ \bibinfo {pages} {1021} (\bibinfo {year}
  {2016})}\BibitemShut {NoStop}%
\bibitem [{\citenamefont {Endres}\ \emph {et~al.}(2016)\citenamefont {Endres},
  \citenamefont {Bernien}, \citenamefont {Keesling}, \citenamefont {Levine},
  \citenamefont {Anschuetz}, \citenamefont {Krajenbrink}, \citenamefont
  {Senko}, \citenamefont {Vuletic}, \citenamefont {Greiner},\ and\
  \citenamefont {Lukin}}]{endres2016atom}%
  \BibitemOpen
  \bibfield  {author} {\bibinfo {author} {\bibfnamefont {M.}~\bibnamefont
  {Endres}}, \bibinfo {author} {\bibfnamefont {H.}~\bibnamefont {Bernien}},
  \bibinfo {author} {\bibfnamefont {A.}~\bibnamefont {Keesling}}, \bibinfo
  {author} {\bibfnamefont {H.}~\bibnamefont {Levine}}, \bibinfo {author}
  {\bibfnamefont {E.~R.}\ \bibnamefont {Anschuetz}}, \bibinfo {author}
  {\bibfnamefont {A.}~\bibnamefont {Krajenbrink}}, \bibinfo {author}
  {\bibfnamefont {C.}~\bibnamefont {Senko}}, \bibinfo {author} {\bibfnamefont
  {V.}~\bibnamefont {Vuletic}}, \bibinfo {author} {\bibfnamefont
  {M.}~\bibnamefont {Greiner}}, \ and\ \bibinfo {author} {\bibfnamefont
  {M.~D.}\ \bibnamefont {Lukin}},\ }\href@noop {} {\bibfield  {journal}
  {\bibinfo  {journal} {Science}\ }\textbf {\bibinfo {volume} {354}},\ \bibinfo
  {pages} {1024} (\bibinfo {year} {2016})}\BibitemShut {NoStop}%
\bibitem [{\citenamefont {Norcia}\ \emph {et~al.}(2018)\citenamefont {Norcia},
  \citenamefont {Young},\ and\ \citenamefont
  {Kaufman}}]{norcia2018microscopic}%
  \BibitemOpen
  \bibfield  {author} {\bibinfo {author} {\bibfnamefont {M.}~\bibnamefont
  {Norcia}}, \bibinfo {author} {\bibfnamefont {A.}~\bibnamefont {Young}}, \
  and\ \bibinfo {author} {\bibfnamefont {A.}~\bibnamefont {Kaufman}},\
  }\href@noop {} {\bibfield  {journal} {\bibinfo  {journal} {Physical Review
  X}\ }\textbf {\bibinfo {volume} {8}},\ \bibinfo {pages} {041054} (\bibinfo
  {year} {2018})}\BibitemShut {NoStop}%
\bibitem [{\citenamefont {Cooper}\ \emph {et~al.}(2018)\citenamefont {Cooper},
  \citenamefont {Covey}, \citenamefont {Madjarov}, \citenamefont {Porsev},
  \citenamefont {Safronova},\ and\ \citenamefont
  {Endres}}]{cooper2018alkaline}%
  \BibitemOpen
  \bibfield  {author} {\bibinfo {author} {\bibfnamefont {A.}~\bibnamefont
  {Cooper}}, \bibinfo {author} {\bibfnamefont {J.~P.}\ \bibnamefont {Covey}},
  \bibinfo {author} {\bibfnamefont {I.~S.}\ \bibnamefont {Madjarov}}, \bibinfo
  {author} {\bibfnamefont {S.~G.}\ \bibnamefont {Porsev}}, \bibinfo {author}
  {\bibfnamefont {M.~S.}\ \bibnamefont {Safronova}}, \ and\ \bibinfo {author}
  {\bibfnamefont {M.}~\bibnamefont {Endres}},\ }\href@noop {} {\bibfield
  {journal} {\bibinfo  {journal} {Physical Review X}\ }\textbf {\bibinfo
  {volume} {8}},\ \bibinfo {pages} {041055} (\bibinfo {year}
  {2018})}\BibitemShut {NoStop}%
\bibitem [{\citenamefont {Saskin}\ \emph {et~al.}(2019)\citenamefont {Saskin},
  \citenamefont {Wilson}, \citenamefont {Grinkemeyer},\ and\ \citenamefont
  {Thompson}}]{saskin2019narrow}%
  \BibitemOpen
  \bibfield  {author} {\bibinfo {author} {\bibfnamefont {S.}~\bibnamefont
  {Saskin}}, \bibinfo {author} {\bibfnamefont {J.}~\bibnamefont {Wilson}},
  \bibinfo {author} {\bibfnamefont {B.}~\bibnamefont {Grinkemeyer}}, \ and\
  \bibinfo {author} {\bibfnamefont {J.~D.}\ \bibnamefont {Thompson}},\
  }\href@noop {} {\bibfield  {journal} {\bibinfo  {journal} {Phys. Rev. Lett.}\
  }\textbf {\bibinfo {volume} {122}},\ \bibinfo {pages} {143002} (\bibinfo
  {year} {2019})}\BibitemShut {NoStop}%
\bibitem [{\citenamefont {van Bijnen}\ and\ \citenamefont
  {Pohl}(2015)}]{vanBijnen15}%
  \BibitemOpen
  \bibfield  {author} {\bibinfo {author} {\bibfnamefont {R.}~\bibnamefont {van
  Bijnen}}\ and\ \bibinfo {author} {\bibfnamefont {T.}~\bibnamefont {Pohl}},\
  }\href {http://journals.aps.org/prl/abstract/10.1103/PhysRevLett.114.243002}
  {\bibfield  {journal} {\bibinfo  {journal} {Phys. Rev. Lett.}\ }\textbf
  {\bibinfo {volume} {114}},\ \bibinfo {pages} {243002} (\bibinfo {year}
  {2015})}\BibitemShut {NoStop}%
\bibitem [{\citenamefont {Glaetzle}\ \emph {et~al.}(2015)\citenamefont
  {Glaetzle}, \citenamefont {Dalmonte}, \citenamefont {Nath}, \citenamefont
  {Gross}, \citenamefont {Bloch},\ and\ \citenamefont {Zoller}}]{Glaetzle15}%
  \BibitemOpen
  \bibfield  {author} {\bibinfo {author} {\bibfnamefont {A.~W.}\ \bibnamefont
  {Glaetzle}}, \bibinfo {author} {\bibfnamefont {M.}~\bibnamefont {Dalmonte}},
  \bibinfo {author} {\bibfnamefont {R.}~\bibnamefont {Nath}}, \bibinfo {author}
  {\bibfnamefont {C.}~\bibnamefont {Gross}}, \bibinfo {author} {\bibfnamefont
  {I.}~\bibnamefont {Bloch}}, \ and\ \bibinfo {author} {\bibfnamefont
  {P.}~\bibnamefont {Zoller}},\ }\href
  {http://journals.aps.org/prl/abstract/10.1103/PhysRevLett.114.173002}
  {\bibfield  {journal} {\bibinfo  {journal} {Phys. Rev. Lett.}\ }\textbf
  {\bibinfo {volume} {114}},\ \bibinfo {pages} {173002} (\bibinfo {year}
  {2015})}\BibitemShut {NoStop}%
\bibitem [{\citenamefont {Goldschmidt}\ \emph {et~al.}(2016)\citenamefont
  {Goldschmidt}, \citenamefont {Boulier}, \citenamefont {Brown}, \citenamefont
  {Koller}, \citenamefont {Young}, \citenamefont {Gorshkov}, \citenamefont
  {Rolston},\ and\ \citenamefont {Porto}}]{Goldschmidt16}%
  \BibitemOpen
  \bibfield  {author} {\bibinfo {author} {\bibfnamefont {E.~A.}\ \bibnamefont
  {Goldschmidt}}, \bibinfo {author} {\bibfnamefont {T.}~\bibnamefont
  {Boulier}}, \bibinfo {author} {\bibfnamefont {R.~C.}\ \bibnamefont {Brown}},
  \bibinfo {author} {\bibfnamefont {S.~B.}\ \bibnamefont {Koller}}, \bibinfo
  {author} {\bibfnamefont {J.~T.}\ \bibnamefont {Young}}, \bibinfo {author}
  {\bibfnamefont {A.~V.}\ \bibnamefont {Gorshkov}}, \bibinfo {author}
  {\bibfnamefont {S.~L.}\ \bibnamefont {Rolston}}, \ and\ \bibinfo {author}
  {\bibfnamefont {J.~V.}\ \bibnamefont {Porto}},\ }\href {\doibase
  10.1103/PhysRevLett.116.113001} {\bibfield  {journal} {\bibinfo  {journal}
  {Phys. Rev. Lett.}\ }\textbf {\bibinfo {volume} {116}},\ \bibinfo {pages}
  {113001} (\bibinfo {year} {2016})}\BibitemShut {NoStop}%
\bibitem [{\citenamefont {Zeiher}\ \emph {et~al.}(2015)\citenamefont {Zeiher},
  \citenamefont {van Bijnen}, \citenamefont {Schau\ss{}}, \citenamefont {Hild},
  \citenamefont {Choi}, \citenamefont {Pohl}, \citenamefont {Bloch},\ and\
  \citenamefont {Gross}}]{Zeiher16}%
  \BibitemOpen
  \bibfield  {author} {\bibinfo {author} {\bibfnamefont {J.}~\bibnamefont
  {Zeiher}}, \bibinfo {author} {\bibfnamefont {R.}~\bibnamefont {van Bijnen}},
  \bibinfo {author} {\bibfnamefont {P.}~\bibnamefont {Schau\ss{}}}, \bibinfo
  {author} {\bibfnamefont {S.}~\bibnamefont {Hild}}, \bibinfo {author}
  {\bibfnamefont {J.-y.}\ \bibnamefont {Choi}}, \bibinfo {author}
  {\bibfnamefont {T.}~\bibnamefont {Pohl}}, \bibinfo {author} {\bibfnamefont
  {I.}~\bibnamefont {Bloch}}, \ and\ \bibinfo {author} {\bibfnamefont
  {C.}~\bibnamefont {Gross}},\ }\href {\doibase 10.1103/PhysRevX.5.031015}
  {\bibfield  {journal} {\bibinfo  {journal} {Phys. Rev. X}\ }\textbf {\bibinfo
  {volume} {5}},\ \bibinfo {pages} {031015} (\bibinfo {year}
  {2015})}\BibitemShut {NoStop}%
\bibitem [{\citenamefont {Zeiher}\ \emph {et~al.}(2017)\citenamefont {Zeiher},
  \citenamefont {Choi}, \citenamefont {Rubio-Abadal}, \citenamefont {Pohl},
  \citenamefont {van Bijnen}, \citenamefont {Bloch},\ and\ \citenamefont
  {Gross}}]{zeiher2017coherent}%
  \BibitemOpen
  \bibfield  {author} {\bibinfo {author} {\bibfnamefont {J.}~\bibnamefont
  {Zeiher}}, \bibinfo {author} {\bibfnamefont {J.-y.}\ \bibnamefont {Choi}},
  \bibinfo {author} {\bibfnamefont {A.}~\bibnamefont {Rubio-Abadal}}, \bibinfo
  {author} {\bibfnamefont {T.}~\bibnamefont {Pohl}}, \bibinfo {author}
  {\bibfnamefont {R.}~\bibnamefont {van Bijnen}}, \bibinfo {author}
  {\bibfnamefont {I.}~\bibnamefont {Bloch}}, \ and\ \bibinfo {author}
  {\bibfnamefont {C.}~\bibnamefont {Gross}},\ }\href@noop {} {\bibfield
  {journal} {\bibinfo  {journal} {Physical Review X}\ }\textbf {\bibinfo
  {volume} {7}},\ \bibinfo {pages} {041063} (\bibinfo {year}
  {2017})}\BibitemShut {NoStop}%
\bibitem [{blo()}]{blockade}%
  \BibitemOpen
  \href@noop {} {}\bibinfo {note} {In the blockage regime, the effective
  interaction strength scales as $ \sim
  \frac{\Omega^4}{8\Delta^3}\frac{1}{1+|r_i-r_j|^6/R_c^6}$, where the
  interaction range $R_c = \left(-C_6/\Delta\right)^{1/6}$ depends on the van
  der Waals coefficient $C_6$ of the Rydberg-Rydberg interaction, $\Omega$ is
  the driving Rabi frequency and $\Delta$ the associated detuning.}\BibitemShut
  {Stop}%
\bibitem [{\citenamefont {Iadecola}\ \emph {et~al.}(2015)\citenamefont
  {Iadecola}, \citenamefont {Santos},\ and\ \citenamefont
  {Chamon}}]{Iadecola15}%
  \BibitemOpen
  \bibfield  {author} {\bibinfo {author} {\bibfnamefont {T.}~\bibnamefont
  {Iadecola}}, \bibinfo {author} {\bibfnamefont {L.~H.}\ \bibnamefont
  {Santos}}, \ and\ \bibinfo {author} {\bibfnamefont {C.}~\bibnamefont
  {Chamon}},\ }\href {\doibase 10.1103/PhysRevB.92.125107} {\bibfield
  {journal} {\bibinfo  {journal} {Phys. Rev. B}\ }\textbf {\bibinfo {volume}
  {92}},\ \bibinfo {pages} {125107} (\bibinfo {year} {2015})}\BibitemShut
  {NoStop}%
\bibitem [{\citenamefont {Potirniche}\ \emph {et~al.}(2017)\citenamefont
  {Potirniche}, \citenamefont {Potter}, \citenamefont {Schleier-Smith},
  \citenamefont {Vishwanath},\ and\ \citenamefont {Yao}}]{Potirniche17}%
  \BibitemOpen
  \bibfield  {author} {\bibinfo {author} {\bibfnamefont {I.-D.}\ \bibnamefont
  {Potirniche}}, \bibinfo {author} {\bibfnamefont {A.~C.}\ \bibnamefont
  {Potter}}, \bibinfo {author} {\bibfnamefont {M.}~\bibnamefont
  {Schleier-Smith}}, \bibinfo {author} {\bibfnamefont {A.}~\bibnamefont
  {Vishwanath}}, \ and\ \bibinfo {author} {\bibfnamefont {N.~Y.}\ \bibnamefont
  {Yao}},\ }\href {\doibase 10.1103/PhysRevLett.119.123601} {\bibfield
  {journal} {\bibinfo  {journal} {Phys. Rev. Lett.}\ }\textbf {\bibinfo
  {volume} {119}},\ \bibinfo {pages} {123601} (\bibinfo {year}
  {2017})}\BibitemShut {NoStop}%
\bibitem [{dri()}]{drivingf}%
  \BibitemOpen
  \href@noop {} {}\bibinfo {note} {Note that any driving function which changes
  sign every half-period will suffice. In an experiment, perhaps the simplest
  approach to realizing such a modulated Ising coupling is to alternate the
  sign of the detuning between the two halves of the drive period.}\BibitemShut
  {Stop}%
\bibitem [{flo()}]{floquetinteractions}%
  \BibitemOpen
  \href@noop {} {}\bibinfo {note} {We note that as written
  Eqn.~\ref{eq:stroboscopicHamiltonian} is dual to free fermions and would thus
  trivially exhibit exact SZMs, even undimerized. In order to avoid this, we
  add interaction terms of the form $V\sum \sigma^x_i \sigma^x_{i+1}$ to the
  time dependent transverse-field Ising Hamiltonian, and propagate this through
  to the effective Hamiltonian before performing our numerical
  simulations.}\BibitemShut {Stop}%
\bibitem [{\citenamefont {Lee}\ \emph {et~al.}(2017)\citenamefont {Lee},
  \citenamefont {Martin}, \citenamefont {Jau}, \citenamefont {Keating},
  \citenamefont {Deutsch},\ and\ \citenamefont {Biedermann}}]{Lee17}%
  \BibitemOpen
  \bibfield  {author} {\bibinfo {author} {\bibfnamefont {J.}~\bibnamefont
  {Lee}}, \bibinfo {author} {\bibfnamefont {M.~J.}\ \bibnamefont {Martin}},
  \bibinfo {author} {\bibfnamefont {Y.-Y.}\ \bibnamefont {Jau}}, \bibinfo
  {author} {\bibfnamefont {T.}~\bibnamefont {Keating}}, \bibinfo {author}
  {\bibfnamefont {I.~H.}\ \bibnamefont {Deutsch}}, \ and\ \bibinfo {author}
  {\bibfnamefont {G.~W.}\ \bibnamefont {Biedermann}},\ }\href {\doibase
  10.1103/PhysRevA.95.041801} {\bibfield  {journal} {\bibinfo  {journal} {Phys.
  Rev. A}\ }\textbf {\bibinfo {volume} {95}},\ \bibinfo {pages} {041801}
  (\bibinfo {year} {2017})}\BibitemShut {NoStop}%
\bibitem [{\citenamefont {Choi}\ \emph {et~al.}(2015)\citenamefont {Choi},
  \citenamefont {Yao}, \citenamefont {Gopalakrishnan},\ and\ \citenamefont
  {Lukin}}]{choi2015quantum}%
  \BibitemOpen
  \bibfield  {author} {\bibinfo {author} {\bibfnamefont {S.}~\bibnamefont
  {Choi}}, \bibinfo {author} {\bibfnamefont {N.~Y.}\ \bibnamefont {Yao}},
  \bibinfo {author} {\bibfnamefont {S.}~\bibnamefont {Gopalakrishnan}}, \ and\
  \bibinfo {author} {\bibfnamefont {M.~D.}\ \bibnamefont {Lukin}},\ }\href@noop
  {} {\bibfield  {journal} {\bibinfo  {journal} {arXiv preprint
  arXiv:1508.06992}\ } (\bibinfo {year} {2015})}\BibitemShut {NoStop}%
\bibitem [{\citenamefont {Parker}\ \emph {et~al.}(2019)\citenamefont {Parker},
  \citenamefont {Vasseur},\ and\ \citenamefont {Scaffidi}}]{Parker19}%
  \BibitemOpen
  \bibfield  {author} {\bibinfo {author} {\bibfnamefont {D.~E.}\ \bibnamefont
  {Parker}}, \bibinfo {author} {\bibfnamefont {R.}~\bibnamefont {Vasseur}}, \
  and\ \bibinfo {author} {\bibfnamefont {T.}~\bibnamefont {Scaffidi}},\ }\href
  {\doibase 10.1103/PhysRevLett.122.240605} {\bibfield  {journal} {\bibinfo
  {journal} {Phys. Rev. Lett.}\ }\textbf {\bibinfo {volume} {122}},\ \bibinfo
  {pages} {240605} (\bibinfo {year} {2019})}\BibitemShut {NoStop}%
\end{thebibliography}%


\begin{thebibliography}{2}%
\makeatletter
\providecommand \@ifxundefined [1]{%
 \@ifx{#1\undefined}
}%
\providecommand \@ifnum [1]{%
 \ifnum #1\expandafter \@firstoftwo
 \else \expandafter \@secondoftwo
 \fi
}%
\providecommand \@ifx [1]{%
 \ifx #1\expandafter \@firstoftwo
 \else \expandafter \@secondoftwo
 \fi
}%
\providecommand \natexlab [1]{#1}%
\providecommand \enquote  [1]{``#1''}%
\providecommand \bibnamefont  [1]{#1}%
\providecommand \bibfnamefont [1]{#1}%
\providecommand \citenamefont [1]{#1}%
\providecommand \href@noop [0]{\@secondoftwo}%
\providecommand \href [0]{\begingroup \@sanitize@url \@href}%
\providecommand \@href[1]{\@@startlink{#1}\@@href}%
\providecommand \@@href[1]{\endgroup#1\@@endlink}%
\providecommand \@sanitize@url [0]{\catcode `\\12\catcode `\$12\catcode
  `\&12\catcode `\#12\catcode `\^12\catcode `\_12\catcode `\%12\relax}%
\providecommand \@@startlink[1]{}%
\providecommand \@@endlink[0]{}%
\providecommand \url  [0]{\begingroup\@sanitize@url \@url }%
\providecommand \@url [1]{\endgroup\@href {#1}{\urlprefix }}%
\providecommand \urlprefix  [0]{URL }%
\providecommand \Eprint [0]{\href }%
\providecommand \doibase [0]{http://dx.doi.org/}%
\providecommand \selectlanguage [0]{\@gobble}%
\providecommand \bibinfo  [0]{\@secondoftwo}%
\providecommand \bibfield  [0]{\@secondoftwo}%
\providecommand \translation [1]{[#1]}%
\providecommand \BibitemOpen [0]{}%
\providecommand \bibitemStop [0]{}%
\providecommand \bibitemNoStop [0]{.\EOS\space}%
\providecommand \EOS [0]{\spacefactor3000\relax}%
\providecommand \BibitemShut  [1]{\csname bibitem#1\endcsname}%
\let\auto@bib@innerbib\@empty
\bibitem [{\citenamefont {Kennedy}\ and\ \citenamefont
  {Tasaki}(1992)}]{Kennedy:92}%
  \BibitemOpen
  \bibfield  {author} {\bibinfo {author} {\bibfnamefont {T.}~\bibnamefont
  {Kennedy}}\ and\ \bibinfo {author} {\bibfnamefont {H.}~\bibnamefont
  {Tasaki}},\ }\href {\doibase 10.1103/PhysRevB.45.304} {\bibfield  {journal}
  {\bibinfo  {journal} {Phys. Rev. B}\ }\textbf {\bibinfo {volume} {45}},\
  \bibinfo {pages} {304} (\bibinfo {year} {1992})}\BibitemShut {NoStop}%
\bibitem [{\citenamefont {Potirniche}\ \emph {et~al.}(2017)\citenamefont
  {Potirniche}, \citenamefont {Potter}, \citenamefont {Schleier-Smith},
  \citenamefont {Vishwanath},\ and\ \citenamefont {Yao}}]{Potirniche17}%
  \BibitemOpen
  \bibfield  {author} {\bibinfo {author} {\bibfnamefont {I.-D.}\ \bibnamefont
  {Potirniche}}, \bibinfo {author} {\bibfnamefont {A.~C.}\ \bibnamefont
  {Potter}}, \bibinfo {author} {\bibfnamefont {M.}~\bibnamefont
  {Schleier-Smith}}, \bibinfo {author} {\bibfnamefont {A.}~\bibnamefont
  {Vishwanath}}, \ and\ \bibinfo {author} {\bibfnamefont {N.~Y.}\ \bibnamefont
  {Yao}},\ }\href {\doibase 10.1103/PhysRevLett.119.123601} {\bibfield
  {journal} {\bibinfo  {journal} {Phys. Rev. Lett.}\ }\textbf {\bibinfo
  {volume} {119}},\ \bibinfo {pages} {123601} (\bibinfo {year}
  {2017})}\BibitemShut {NoStop}%
\end{thebibliography}%
\bibliographystyle{apsrev4-1}

\end{document}


\title{Supplemental Material for Symmetry enhanced boundary qubits at infinite temperature}

\author{Jack Kemp}
\author{Norman Y. Yao}
\author{Chris R. Laumann}

\maketitle

\noindent {\bf Duality Transformation of ZXZ model}---We describe the duality transformation that takes the 1D Haldane-type $\mathbb{Z}_2 \times \mathbb{Z}_2$ SPT of Eqn.~\ref{main-eq:ZXZ} in the maintext to the global symmetry breaking phase of two coupled Ising chains. This transformation generalizes the non-local unitary described by Kennedy and Tasaki \cite{Kennedy:92}, to reveal the hidden ‘string’ order of the AKLT model. We will derive it first for the ZXZ model of the $\mathbb{Z}_2 \times \mathbb{Z}_2$ SPT phase and then comment on how it generalizes to the full Hamiltonian.
The ZXZ model of the $\mathbb{Z}_2 \times \mathbb{Z}_2$ SPT is given in the commuting limit by the Hamiltonian:
\begin{equation} H_{\textrm{ZXZ}} = \sum_{j=1}^{L-2} \sigma^z_{j} \sigma^x_{j+1}\sigma^z_{j+2}\end{equation}

Pairs of spins that can be arranged into two-site unit cells on this open one dimensional chain, with $L = 2M$ spins in total. The model satisfies the two $\mathbb{Z}_2$ symmetries:
\begin{align} G_e &= \prod_{j=1}^{M} \sigma^x_{2j} \\
G_o &= \prod_{j=1}^{M} \sigma^x_{2j-1}
\end{align}

We see that in this version of the SPT, we may view the $\mathbb{Z}_2 \times \mathbb{Z}_2$  ‘on-site’ symmetry as being defined on the unit cells rather than the individual spins.

We now define the duality transformation U:
\begin{align*}
W &= \prod_{j \text{odd}} (-\sigma^x_{2j}) \\
V &= \prod_j (-\sigma^x_{2j-1} P_j + (1- P_j)) \\
P_j &= \frac{1- \prod_{i<j} \sigma^x_{2i}}{2} \\
U &= VW. \end{align*}

All of these operators commute as they are formed only
out of the $\sigma^x_i$ operators. It can be easily verified that $U$, $V$ and $W$ are unitary and Hermitian, which the $P_j$ are projectors; that is $P_j = P_j^2 = P^\dagger$.

In order to determine the action of $U$ on local operators, we need only determine its action on an operator basis for the chain. This is conveniently provided by the Pauli operators $\sigma^x_i$ and $\sigma^z_i$ on each site $i$, which are respectively symmetry even/odd under the relevant on-site symmetry:
\begin{align*}
G_{e/o} \sigma^x_i G^\dagger_{e/o} &=  \sigma^x_i \\
G_{e/o}  \sigma^z_{2j} G^\dagger_{e/o} &= \mp  \sigma^z_{2j} \\
G_{e/o}  \sigma^z_{2j-1} G^\dagger_{e/o} &= \pm  \sigma^z_{2j-1} \\
\end{align*}

Since $U$ only involves $\sigma^x$ terms, it is clear that $\sigma^x_j$ commutes with $U$, and so that:

\begin{equation}
\sigma^x_i \xrightarrow{U} U^\dagger \sigma^x_i U = \sigma^x_i \label{eq:Xtrans}
\end{equation}

For the odd site $\sigma^z$ operators:
\begin{align}
\sigma^z_{2j-1} &\xrightarrow{U} U^\dagger \sigma^z_{2j-1} U \nonumber \\
& = \left ( \prod_{k <j} \sigma^x_{2k} \right) \sigma^z_{2j-1} \nonumber \\
&= \mathcal{L}_j \sigma^z_{2j-1}, \label{eq:Zetrans} \end{align} where we have defined the left-going string operator $\mathcal{L}_j$ which measures the even-site Ising parity to the left of $j$.

Meanwhile for even site operators we find similarly that:
\begin{align}
\sigma^z_{2j} &\xrightarrow{U} U^\dagger \sigma^z_{2j} U \nonumber \\
& = (-1)^j \sigma^z_{2j} \left ( \prod_{k > j} -\sigma^x_{2k} \right) \nonumber \\
&= (-1)^{j+(M-j-1)} \sigma^z_{2j} \mathcal{R}_j, \label{eq:Zotrans} \end{align} 
where $\mathcal{R}_j =   \prod_{k > j} \sigma^x_{2k-1} $ is the right-going string measuring Ising parity.

From equations \eqref{eq:Xtrans}, \eqref{eq:Zetrans} and \eqref{eq:Zotrans}, we can show that any locally supported operator which is even under both symmetries is mapped to a local operator under $U$. In particular, any strings generated by symmetry odd operators on different sites cancel. We also note that the duality squares to one, justifying the use of the term.

Now we have introduced the duality, we can examine its effect on the Hamiltonian in equation 1 in the maintext. Notice in particular that: \begin{align*}
\sigma^z_{2j} \sigma^x_{2j+1} \sigma^z_{2j+2} &\xleftrightarrow{U} \sigma^z_{2j} \sigma^z_{2j+2} \\
\sigma^z_{2j} \sigma^x_{2j+1} \sigma^z_{2j+2} &\xleftrightarrow{U} \sigma^z_{2j} \sigma^z_{2j+2} \\
 \sigma^x_{i}&\xleftrightarrow{U} \sigma^x_{i}.\end{align*}

Using these relations it is trivial to see that $H_{\textrm{SPT}}$ from equation \eqref{main-eq:ZXZ} in the main text and $H^\prime_{\textrm{SPT}}$ are indeed related by the duality transformation: $H_{\textrm{SPT}} \xleftrightarrow{U} H^\prime_{\textrm{SPT}}$.

Furthermore, under the duality the SPT edge modes, main-text equation \eqref{main-eq:SPTedge}, become:
\begin{align*}
\Sigma^x &= \sigma^x_1 \sigma^z_2 \xleftrightarrow{U} \sigma^x_1 \sigma^z_2 \mathcal{R}_1 = \sigma^z_2 G_o \nonumber\\
\Sigma^y &= \sigma^y_1 \sigma^z_2 \xleftrightarrow{U} = i \sigma^z_1 \sigma^z_2 G_o  \nonumber\\
\Sigma^z &= \sigma^z_1 \xleftrightarrow{U} \sigma^z_1  \end{align*}
So, given that $G_o$ is conserved, the autocorrelation times of the SPT edge modes directly follow from the lifetime of the edge magnetisation on each Ising chain.

{\bf Numerical Simulation of Floquet Hamiltonian }---We simulate the effective Floquet Hamiltonian which governs our proposed experimental implementation in the main text using exact diagonalisation for system sizes up to 14 sites long. The Hamiltonian with interactions is given by:
\begin{align} \label{eq:floquet}
H^{(0)}_{\mathrm{F}} = \sum_{i=1}^{N}& h_{i}a(\lambda_{1}, \lambda_{2}) \sigma_{i}^{x} - \sum_{i=2}^{N-1}h_{i}b(\lambda_{1}, \lambda_{2}) \sigma_{i-1}^{z}\sigma_{i}^{x}\sigma_{i+1}^{z}\\ \nonumber
&+ V_{x}c_{\rm edge}(\lambda_{2})(\sigma_{1}^{x}\sigma_{2}^{x} + \sigma_{N-1}^{x}\sigma_{N}^{x})\\ \nonumber
&+ V_{x}\sum_{i=2}^{N-2}\left[c(\lambda_{i+1})\sigma_{i}^{x}\sigma_{i+1}^{x} +d(\lambda_{i+1})\sigma_{i-1}^{z}\sigma_{i}^{y}\sigma_{i+1}^{y}\sigma_{i+2}^{z}\right], \nonumber
\end{align}
where $J_{0}(x)$ is the Bessel function of the first kind, $a(\lambda_{1}, \lambda_{2}) = \frac{1}{2} \left[J_{0}\left(2(\lambda_{1} - \lambda_{2})\right) + J_{0}(2(\lambda_{1}+\lambda_{2})) \right]$,  $b(\lambda_{1},\lambda_{2})  = J_{0}(2(\lambda_{1}-\lambda_{2})) - a(\lambda_{1}, \lambda_{2})$, $c(\lambda)=\frac{1}{2}\left[1+J_{0}(4\lambda)\right]$,  $d(\lambda)=1-c(\lambda)$, and $c_{\rm edge}(\lambda) = J_{0}(2\lambda)$~\cite{Potirniche17}.

\begin{figure}[h!]
 \includegraphics[width=0.6\columnwidth]{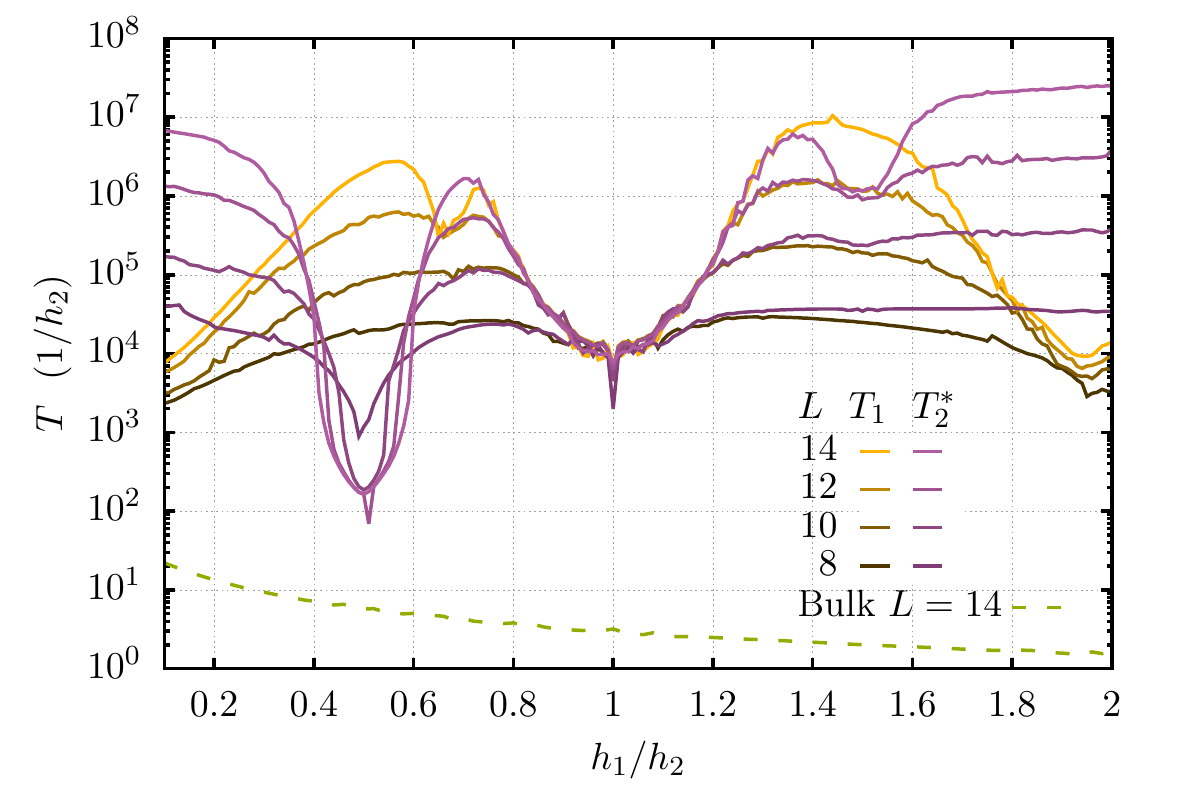}
\caption{The decay times $T_1$ and $T_2^*$ of the autocorrelators of the conjugate edge operators $\Sigma^z$ and $\Sigma^x$ respectively for the effective Floquet Hamiltonian, Eq.~\eqref{eq:floquet}, compared to a bulk spin $\sigma^z_7$.
These results are calculated using exact diagonalisation at infinite temperature with $V = b/a = 0.05$, and $\lambda_1 = 2.68$ and $\lambda_2 = 1.20$. Despite the resonance structure, edge enhancement of coherence times over the bulk is clearly visible.}
\label{fig:varyh1}
\end{figure}

\begin{figure}[h!]
 \includegraphics[width=0.6\columnwidth]{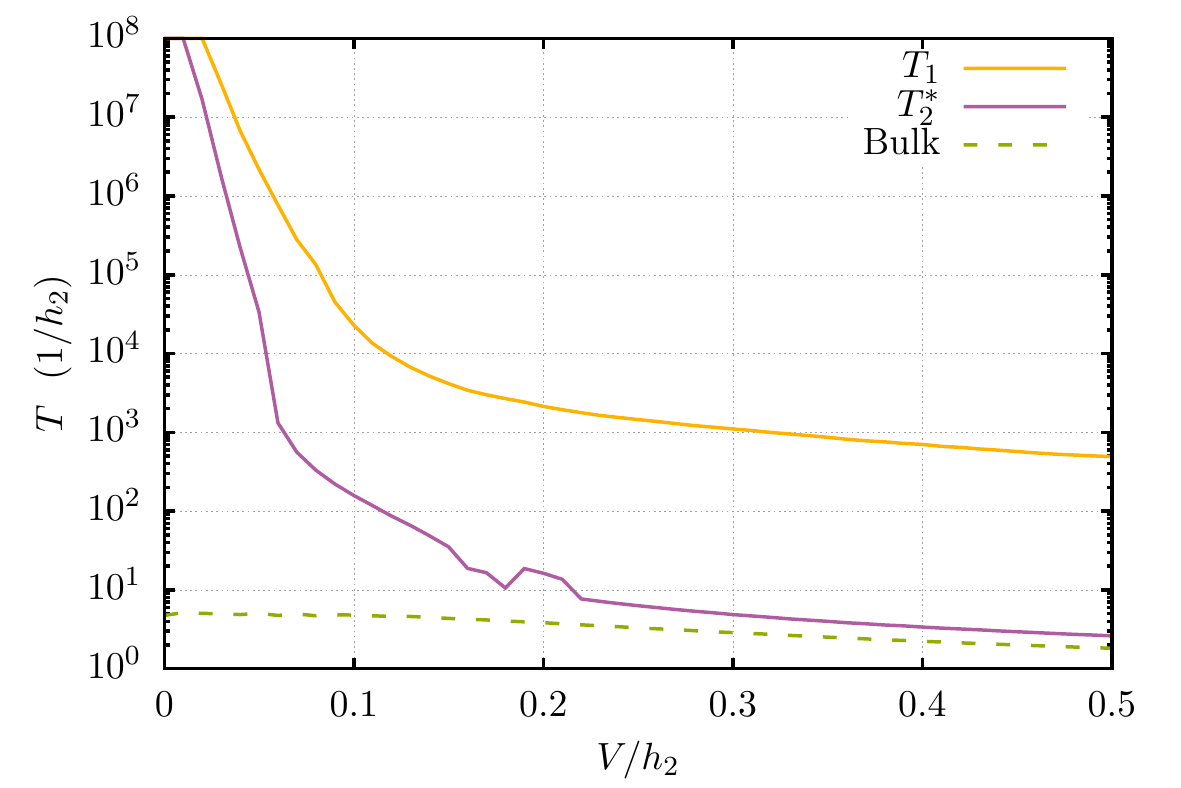}
\caption{The decay times $T_1$ and $T_2^*$ of the autocorrelators of the conjugate edge operators $\Sigma^z$ and $\Sigma^x$ respectively for the effective Floquet Hamiltonian, Eq.~\eqref{eq:floquet} at system size $L=14$, compared to a bulk spin $\sigma^z_7$.
These results are calculated using exact diagonalisation at infinite temperature with $V = b/a = 0.05$, and $\lambda_1 = 2.68$ and $\lambda_2 = 1.20$.}
\label{fig:varyV}
\end{figure}

In order to compare with the numerics in the main text we choose $\lambda_1 = 2.68$ and $\lambda_2 = 1.20$ such that $a/b = 0.05$. This choice also ensures that $c_{\rm edge}$ is negligible. Figure~\ref{fig:varyh1} shows the autocorrelation time of the edge modes as $h_1/h_2$ is varied. A similar resonance structure is observed as for the ZXZ model in the main text, with the exception that the resonance at the swap-symmetric point is no longer a first order resonance as the term driving the relevant domain-wall conversion, $\sigma^x_1 \sigma^x_2$, has dropped out of the Hamiltonian with $c_{\rm edge}$. The edge enhancement of the autocorrelation times relative to the bulk is dramatic. In Figure~\ref{fig:varyV} the interaction strength $V$ is instead varied at $h_1/h_2 = 0.6$. For $V \lesssim 0.2 h_2$ both $T_1$ and $T_2^*$ enjoy protection at the edge. For interaction strengths greater than this the resonance at $h_1 = h_2/2$ for $\Sigma^x$ becomes wide enough to completely suppress its autocorrelation time. Nevertheless, it is clear that there is a wide range of parameters for which edge enhancement of the autocorrelation times relative to the bulk should be observed.

 \bibliography{szm}
\bibliographystyle{apsrev4-1}